 \documentclass[preprint,journal]{vgtc}            

\usepackage{xspace}

\onlineid{2104}

\vgtccategory{Research}

\title{\ournamenospace: A Stage-Adaptive Immersive Visual Analytics Framework for Anatomical Liver Surgical Planning}

\author{%
  \authororcid{Qixuan Liu}{0009-0001-5343-8052},
  \authororcid{Shi Qiu}{0000-0001-9958-180X},
  Xiwen Wu,
  Yuqi Tong,
  \authororcid{Yinqiao Wang}{0000-0002-6099-206X},
  \authororcid{Ruiyang Li}{0000-0003-3880-068X},
  \authororcid{Jialun Pei}{0000-0002-2630-2838},\\
  \authororcid{Shengdong Zhao}{0000-0001-7971-3107},
  \authororcid{Chi-Wing Fu}{0000-0002-5238-593X}, and
  \authororcid{Pheng-Ann Heng}{0000-0003-3055-5034}
}

\authorfooter{
  \item
  	Qixuan Liu, Shi Qiu, Yuqi Tong, Yinqiao Wang, Ruiyang Li, Jialun Pei, Chi-Wing Fu, Pheng-Ann Heng are with Department of Computer Science and Engineering and Institute of Medical Intelligence and XR, The Chinese University of Hong Kong.
  	E-mails: \{qxliu, shiqiu, yqtong, yqwang, liry, cwfu, pheng\}@cse.cuhk.edu.hk, jialunpei@cuhk.edu.hk.
  \item
  	Shengdong Zhao is with School of Creative Media, City University of Hong Kong.
  	E-mail: shengdong.zhao@cityu.edu.hk.
  \item Xiwen Wu is with First Department of Hepatobiliary Surgery, Zhujiang Hospital, Southern Medical University.
  	E-mail: ngheiman@126.com.
  \item Corresponding author: Shi Qiu. 
}

\abstract{Anatomical liver resection (ALR) surgery is the most important treatment for liver cancer, yet preoperative planning demands complex, multi-stage clinical reasoning under competing safety constraints.
Current 2D desktop tools are not well equipped to support this process, exhibiting three fundamental limitations: reliance on monolithic interfaces that fail to adapt to the distinct cognitive demands of each planning stage; a perceptual bottleneck caused by limited anatomical spatial representation and missing plane-vessel intersection visualization; and an attention bottleneck stemming from fragmented critical safety criteria display across separate views.
We present \ourname, a stage-adaptive immersive visual analytics framework for ALR planning, grounded in an 8-month collaboration with two expert hepatobiliary surgeons.
Decomposing the surgical planning process into three sequential yet cognitively distinct stages, \ourname externalizes the cognitive demand of each stage via tailored techniques:
(1) context-preserving focus and hue-preserving rendering for anatomical discovery; (2) direct 3D resection plane manipulation coupled with real-time, embedded visual feedback on critical safety criteria during plan refinement; and (3) explicit plane-vessel intersection visualization for anticipatory surgery preparation.
A within-subjects study with eight hepatobiliary surgeons against a \revrep{desktop baseline}{desktop-based system} shows large-effect-size improvements in task completion time, perceived cognitive workload, and system usability \revadd{on controlled planning tasks}.
Moreover, our \revrep{study}{evaluation} reveals broader insights: \revrep{\ourname}{stage-adaptive design} reduces cognitive burden and \revrep{encourages a shift in}{shifts} surgeons from merely satisfying safety criteria to actively optimizing them, \revrep{suggesting that}{while} explicit visualization of spatial relationships lowers the cognitive barrier to \revrep{complex}{reliable} surgical planning.}


\usepackage{mathptmx}                  
\usepackage{xspace}
\usepackage{soul}
\usepackage{booktabs}
\usepackage{multirow} 
\usepackage{threeparttable}
\usepackage{amssymb}
\usepackage{xurl}
\newif\ifshowrevisions
\ifshowrevisions
  \usepackage[authormarkup=none]{changes}
\else
  \usepackage[final]{changes}
\fi

\newcommand{\ournamenospace}{\textsc{LiverPlan}}
\newcommand{\ourname}{\textsc{LiverPlan}\xspace} 
\definechangesauthor[name={Revision}, color=red]{R}

\newcommand{\revadd}[1]{\added[id=R]{#1}}
\newcommand{\revdel}[1]{\deleted[id=R]{#1}}
\newcommand{\revrep}[2]{\replaced[id=R]{#1}{#2}}

\newcommand{\eg}{{\em{e.g.}}\@\xspace}

\begin{document}


\maketitle

\section{Introduction}

According to the WHO-affiliated GLOBOCAN 2022 report~\cite{bray2024global}, liver cancer is among the top three leading causes of cancer mortality worldwide.
To effectively treat liver cancer, anatomical liver resection (ALR) is widely recognized as a cornerstone surgical approach that aims at completely removing liver tissues, considering specific vessel territories while preserving sufficient liver functionality~\cite{makuuchi1985ultrasonically, hasegawa2005prognostic, huang2017meta}.

Successful ALR planning requires simultaneous satisfaction of three critical safety criteria~\cite{galle2018easl}: 
(i) sufficient liver volume remains to prevent postoperative liver failure (\textbf{Functional Liver Remnant, FLR}~\cite{memeo2021optimization}); 
(ii) an adequate safety distance is maintained between the tumor boundary and each resection plane (\textbf{Resection Margin, RM}~\cite{lin2022prognostic}); and 
(iii) the resection volume preferably encompasses all tumor-affected segments (\textbf{Target Segment Complete Removal, TSCR}~\cite{bismuth1982surgical}).
These criteria present inherent trade-offs,~\eg, enlarging the resection volume to satisfy TSCR and RM reduces FLR, whereas conservative cuts risk incomplete tumor removal.
Resolving this multi-constraint spatial trade-off requires complex, multi-stage 3D spatial reasoning on a case-by-case basis, where existing planning tools are \revrep{usually}{clearly} inadequate.

Effective ALR preoperative planning follows a structured, three-stage cognitive progression: 
(i) anatomical analysis to identify target segments, 
(ii) iterative resection plan construction against critical safety criteria, and 
(iii) spatial examination of plane-vessel intersections for surgery preparation.
Yet, standard 2D tools exhibit three fundamental limitations across these stages.
First, they rely on \textit{monolithic interfaces} and static visual mappings, failing to adapt to the distinct cognitive demands of each planning stage.
Second, a \textit{perceptual bottleneck} impairs both anatomical analysis and surgery preparation. 
The lack of context-preserving focus leaves internal structures severely occluded by dense anatomical volume meshes, preventing reliable analysis of spatial relationships. 
Moreover, the absence of explicit plane-vessel intersection visualization forces surgeons to rely heavily on internal working memory to mentally reconstruct the intersection geometry.
Third, an \textit{attention bottleneck} disrupts interactive plan refinement. 
Relying on abstract 2D control-point interfaces makes 3D plane manipulation laborious. 
Additionally, monitoring critical safety criteria (FLR, RM, TSCR) in spatially separate panels forces repeated attention-switching, increasing working memory load and the risk of safety oversight.

To address these limitations, we collaborate closely with two experienced hepatobiliary surgeons to design \ournamenospace, a stage-adaptive immersive visual analytics framework for ALR preoperative planning.
\ourname leverages the spatial tracking and immersive display capabilities of mobile extended reality (XR) headsets to realize \textit{stage-adaptive} visualization and interaction.
\revadd{The XR environment supports stereoscopic perception, alleviating the \textit{perceptual bottleneck} in spatial understanding. It also enables 6-DoF interaction for direct 3D plane manipulation, reducing the interaction-related \textit{attention bottleneck} and aligning with the spatial reasoning demands of preoperative ALR planning.}
Importantly, we articulate the ALR planning progression into a three-stage visual analytics workflow and design dedicated stage-adaptive visual encodings and interaction mechanisms to meet specific needs and offload cognitive demand~\cite{risko2016cognitive}: 
(i) Stage 1: facilitating anatomical discovery by resolving severe visual occlusion \revrep{via}{through} context-preserving focus and maintaining color fidelity; 
(ii) Stage 2: streamlining multi-constraint planning via direct 3D manipulation \revdel{coupled} with real-time, embedded visual feedback on critical safety criteria (FLR\revrep{/}{, }RM\revrep{/}{, }TSCR); 
and 
(iii) Stage 3: supporting anticipatory mental model construction \revrep{via}{through} explicit plane-vessel intersection visualization.

The main contributions of this paper are summarized as follows.
\textbf{(i) A cognitively-grounded visual analytics workflow} for ALR surgical planning, derived from an 8-month longitudinal collaboration with hepatobiliary surgeons.
We characterize the planning process as three cognitively distinct visualization stages, each formalized with abstract analytical tasks, design goals, and design requirements.
This evidence-based decomposition provides a principled design framework to inform future stage-adaptive surgical visualization systems.
\textbf{(ii) Stage-adaptive visualization and interaction techniques} addressing stage-specific perceptual and cognitive demands: context-preserving focus and hue-preserving rendering strategy to externalize spatial perception for anatomical discovery; direct 3D manipulation with real-time, embedded safety criteria visualization (color-coded FLR remnant outline, dual-mode RM distance feedback, and TSCR intersection contours) to externalize constraint status during resection plane manipulation; and explicit plane-vessel intersection visualization to support anticipatory mental model construction.
We implement these techniques as a working XR prototype deployed on a commodity headset (Meta Quest 3, \cref{fig:system_interface}).
\textbf{(iii) Empirical evaluation with eight hepatobiliary surgeons}, demonstrating significant \revadd{overall} improvements in task completion time, perceived cognitive workload, and system usability over \revrep{a desktop baseline on controlled planning tasks.}{desktop planning tools, with validated stage-specific effectiveness.}
\section{Related Work}

Recent advances in computer-aided surgical planning aim at 
improving precision and safety in anatomical liver resection (ALR).
Complementary research areas, including visual analytics for clinical decision-making, immersive medical visualization, and embedded visualization, provide valuable strategies for externalizing spatial reasoning and streamlining multi-constraint 
workflows.
We position this work in these research, addressing existing research gaps in these fields.

\subsection{Visual Analytics for Clinical Decision Making}
Decision support visualizations are important for complex data exploration.
By transforming the exploratory process into actionable visual insights, human reasoning and judgment can be largely enhanced~\cite{thomas2005illuminating, keim2008visual}.
In clinical domains, purely automated models or black-box predictions are often inadequate for accurate decision making~\cite{quinn2022three}. 
This emphasizes the critical role of visual analytics in 
keeping domain experts in the loop to derive robust, informed decisions~\cite{assadi2022decision, knittel2025embryoprofiler}.

Within the broader area of clinical decision support, structural surgical planning (\eg, ALR planning) represents a highly specialized subset that demands precise 3D spatial reasoning and multi-constraint trade-offs. 
To support this computationally demanding process, various computer-aided systems have been developed to enhance preoperative planning through dedicated visualizations~\cite{fedorov20123d, materialise_mimics, reitinger2006liver, boedecker2021using, reinschluessel2022virtual}. 
However, deploying effective visual support for complex tasks requires strict alignment with established medical practices and cognitive patterns, which often necessitates close collaboration with practitioners through workflow-centric~\cite{sedlmair2012design, bhat2023towards} and decision-centered designs~\cite{militello201315}.

In cognitive design and visual analytics, researchers advocate for progressive disclosure~\cite{spillers2004progressive} and progressive visual analytics paradigms~\cite{stolper2014progressive} to reduce load by structuring overwhelming analytical processes into manageable, sequential stages.
Despite this, state-of-the-art computer-aided surgical tools predominantly treat spatial planning as a single task, employing monolithic interfaces across the entire workflow~\cite{fedorov20123d, materialise_mimics, boedecker2021using, reinschluessel2022virtual}. 
Our work bridges this methodological gap by bringing visual analytics principles into surgical planning: we characterize the implicit, multi-stage planning workflow as structured analytical tasks, providing an evidence-based framework that integrates stage-adaptive visual analytics directly into the surgical context.

\subsection{Immersive Medical Visualization}

\revdel{Anatomical medical data are inherently three-dimensional.}
Exploring 3D anatomy through 2D desktop interfaces introduces perceptual bottlenecks.
Users must mentally reconstruct spatial structure from cross-sectional slices, and transparency-based compositing on flat screens compromises the depth and color cues essential for tumor-vessel assessment~\cite{andriole2011optimizing}. 
To overcome these bottlenecks, immersive VR/AR/XR technologies are increasingly adopted to offer intuitive depth perception and six-DoF spatial reasoning,~\eg, 
enhancing surgeons' ability to comprehend complex spatial relationships in locating tumors relative to critical vasculature~\cite{hattab2021investigating, yuan2023extended, venkatesan2021virtual, jadhav2022md} and cone-based or free-form \revrep{resection}{resurgical planning section} visualization~\cite{chheang2021collaborative, reinschluessel2022virtual}.
However, existing immersive applications largely implement static visual mappings across the entire application lifecycle, focusing on generic visualization tools \revrep{to render}{for rendering the} 3D anatomy.
They lack adaptive configurations to support the shifting perceptual focuses \revrep{in}{during} the clinical procedure.
\revrep{\ourname aligns}{In this work, aligning} with the principles of \textit{immersive analytics}~\cite{marriott2018immersive, fonnet2019survey}, \revrep{leveraging}{\ourname leverages} the XR medium as a responsive visual analytic space, \revrep{dynamically adapting}{where} visualization strategies \revdel{are dynamically adapted} to the cognitive requirements of each planning stage.

\subsection{Embedded Visualization}

A persistent challenge in multi-constraint surgical planning is the attention bottleneck caused by information fragmentation.
Standard interfaces spatially decouple the 3D anatomical view from associated numerical dashboards monitoring vital constraints,~\eg, the resection margin and the remnant volume in our task.
Cognitive psychology and visualization design guidelines underscore that such spatial separation obligates users to frequently split their attention, substantially increasing working memory load and cognitive cost~\cite{wickens2008multiple, Sweller2011, baldonado2000guidelines, shao2020teaching}.
To mitigate this divided attention~\cite{chandler1992split}, the framework of embedded data representations~\cite{willett2017embedded} advocates for tightly coupling data visualizations with their physical or conceptual referents. 
In surgical and healthcare contexts, integrated visual feedback has proven effective at keeping contextual information within the primary focus area~\cite{horeman2012visual, tu2024head}.
While immersive analytics systems are actively exploring embedded visualizations, these efforts predominantly focus on affixing static semantic labels or charts to 3D objects~\cite{tatzgern2014hedgehog, lin2021labeling, ens2021grand}. 
By embedding continuous multi-constraint feedback directly in the focused area, we eliminate off-target visual saccades, demonstrating how embedded visualization can uniquely support interactive surgical planning.

\revrep{Yet}{However}, realizing embedded visualizations in 3D surgical environments presents unique design challenges.
It requires real-time encoding of multi-constraint safety metrics \revrep{in}{within} a 3D focused area, without causing visual clutter or occluding underlying anatomy. 
To our best knowledge, \ourname presents \revrep{a new}{the first} immersive visual analytics system to \revrep{address}{overcome} these challenges, explicitly designed to support ALR planning \revrep{via}{through} our stage-adaptive workflow and real-time embedded visualizations.
\section{Clinical Background}\label{sec:clinical_bg}

\begin{figure}[tb]
  \centering
  \includegraphics[width=\linewidth]{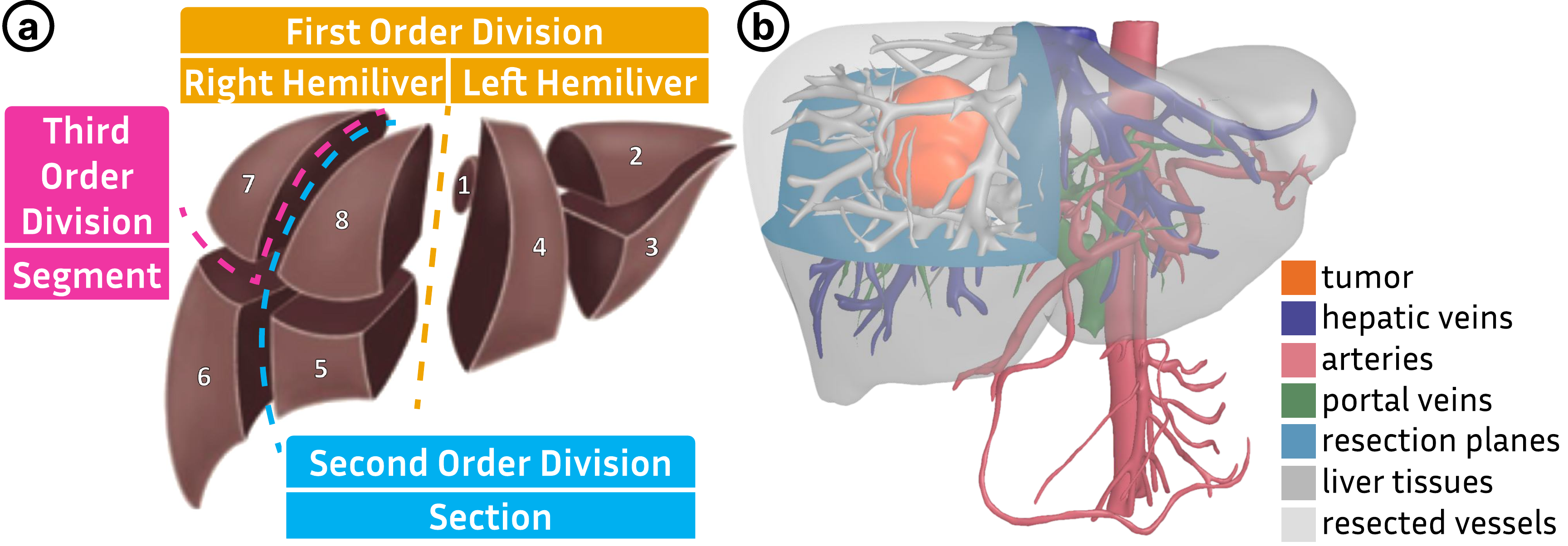}
  \caption{Liver in anatomical resection. (a) illustrates the hierarchical clinical division of the liver: 1st-order division into hemilivers, 2nd-order into sections, and 3rd-order into the eight Couinaud segments~\cite{couinaud1957foie}. \textit{This sub-figure is reproduced from} Faria \emph{et al.}~\cite{faria2022liver}. (b) shows the internal structures: main vessels (arteries in red, hepatic veins in blue, portal veins in green), the tumor (orange), and example resection planes (light blue) used for target segments removal.
  }
  \label{fig:liver_anatomy}
\end{figure}
 
Fig.~\ref{fig:liver_anatomy} (a) illustrates the anatomical structure of the liver, which is hierarchically divided into eight Couinaud segments based on vascular and biliary territories~\cite{couinaud1957foie, strasberg2000brisbane}:
1st-order division into hemilivers (segments 1,2,3,4 and 5,6,7,8), 2nd-order division into sections (segments 6,7 and 5,8), and 3rd-order division into the eight Couinaud segments.

\begin{figure*}[tb]
  \centering
  \includegraphics[width=\linewidth]{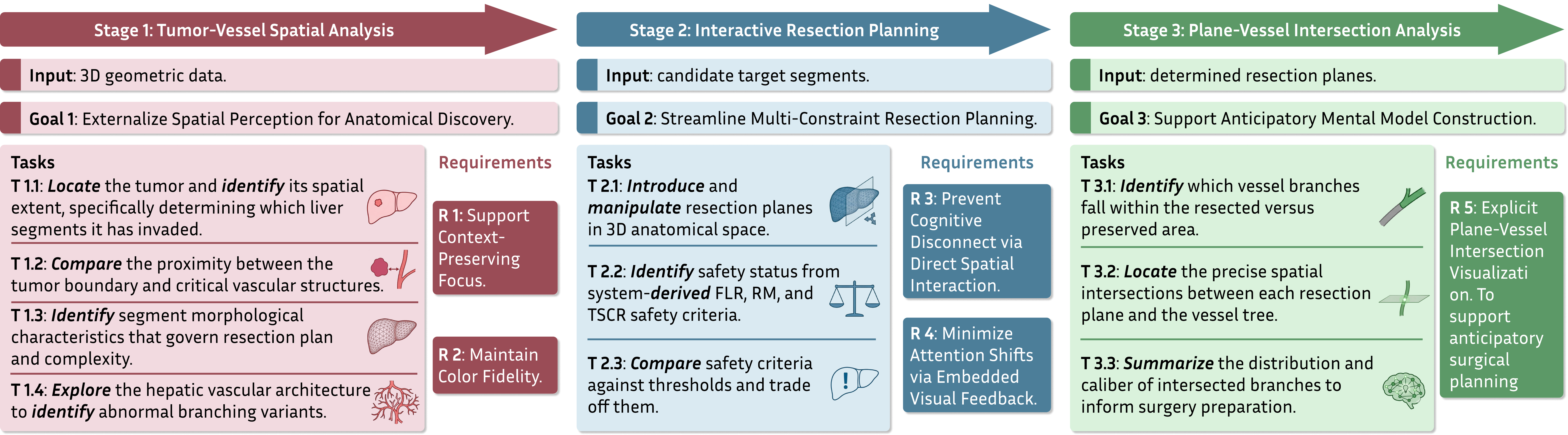}
  \caption{The cognitively-grounded ALR planning workflow diagram.
  ALR planning is decomposed into three sequential, distinct cognitive stages structured according to the Multi-Level Typology of Abstract Visualization Tasks~\cite{brehmer2013multi}.
  Each stage (red, blue, green) specifies its \textbf{Input}, analytical \textbf{Goal}, domain \textbf{Tasks} (\textbf{T}), and the derived design \textbf{Requirements} (\textbf{R}).
  }
  \label{fig:cog_workflow}
\end{figure*}

A key strategy in liver resection surgery is Anatomical Liver Resection (ALR)~\cite{makuuchi1985ultrasonically, hasegawa2005prognostic, huang2017meta}.
It aims at segment-level removal, to resect entire Couinaud segment(s) that contain tumors to ensure clearance of small cancer cells~\cite{nakashima1986pathologic, imamura1999prognostic, faria2022liver}. 
As illustrated in Fig.~\ref{fig:liver_anatomy} (b), ALR planning centers on the interplay among three key anatomical elements within the liver: the vessels (arteries, hepatic veins, and portal veins), the tumor, and the resection planes.
Vessels supplying the resected segment are subsequently ligated, while those within the preserved remnant must be protected.
Successful surgical planning for ALR has to satisfy three critical safety criteria~\cite{galle2018easl}:
(i) \textbf{Functional Liver Remnant (FLR)} avoids excessive removal of the liver,~\emph{i.e.}, preserving sufficient functional remnant liver volume (typically > 40\% of total liver volume)~\cite{memeo2021optimization}; 
(ii) \textbf{Resection Margin (RM)} suggests maintaining an adequate minimum distance (typically > 1cm) between the tumor boundary and resection plane, to avoid missing 
potential cancer cells~\cite{lin2022prognostic}; and 
(iii) \textbf{Target Segment Complete Removal (TSCR)} suggests completely removing target liver segment(s) to prevent cancer recurrence~\cite{makuuchi1985ultrasonically, hasegawa2005prognostic, huang2017meta, bismuth1982surgical}.
%
In practice, trade-offs between these criteria are often necessary, with FLR prioritized over RM and RM over TSCR.
\section{Design Goals and Requirements}\label{sec:design_goals}


To unveil the domain challenges and derive our visualization designs, we take a user-centered design approach through an eight-month close collaboration with two hepatobiliary surgeons: 
an attending surgeon (S1, male, 18 years of experience) and a resident surgeon (S2, female, 5 years of experience).
\revadd{All studies involving surgeon participants reported in this paper---including the formative study, case study, and user study---were approved by the Joint CUHK--NTEC Clinical Research Ethics Committee (CREC; Ref.\ No.\ 2022.142), and informed consent was obtained from all participants.}

The process comprised two phases:
(i) \textit{Clinical Discovery (12 weeks)} and (ii) \textit{Iterative Design \& Refinement (20 weeks)}. We first sought to ground our understanding in actual clinical workflows. Through 11 bi-weekly interviews (15--40 mins each) with both surgeons, participation in four surgical planning seminars, and observation of a three-hour live liver resection surgery, we analyzed standard-of-care practices and documented the visualization bottlenecks inherent to current 2D desktop planning tools.
Based on our initial findings, we then continuously develop and refine our visual analytics framework. This was driven by 11 bi-weekly design sessions with S2, supplemented by 15 weekly ad-hoc consultations. To ensure clinical validity and strategic alignment, we conducted monthly validation meetings (3 sessions) with the senior surgeon (S1). This rigorous multidisciplinary collaboration directly informed the workflow characterization below.

\subsection{Cognitively-Grounded Visual Analytics Workflow}\label{subsec:workflow}



Standard 2D planning tools provide a single monolithic interface for the entire ALR procedure, imposing compound cognitive burdens that fail to align with the highly structured, multi-stage nature of surgical reasoning.
To systematically formalize this multi-stage surgical reasoning into actionable visualization tasks, we synthesize the domain findings from our collaborative design process using the \textbf{Multi-Level Typology of Abstract Visualization Tasks}~\cite{brehmer2013multi}.
This framework resolves the ambiguity between the ``ends'' and ``means'' of analytical tasks by characterizing user behavior along three standardized dimensions: \textit{why} a task is performed (spanning from high-level goals like \textit{discover} to low-level queries like \textit{identify} or \textit{compare}), \textit{how} it is executed using specific visualization methods (\eg, \textit{navigate} or \textit{filter}), and \textit{what} inputs and outputs the task pertains to.

Applying this typology to our clinical observations, we identify that ALR planning naturally separates into three distinct cognitive phases, each characterized by different analytical objectives, information dependencies, and attention demands.
Because this decomposition emerges from observed surgical cognition and was iteratively validated by both collaborating surgeons (S1, S2), it provides a principled, cognitively grounded, and expert-confirmed basis for expressing complex surgical planning as sequences of interdependent tasks.
In the detailed stage descriptions below, we \emph{italicize} the formal terms defined in~\cite{brehmer2013multi} to explicitly map the surgeons' domain-specific actions to the standardized visual analytics terms.
\cref{fig:cog_workflow} provides a structured overview of this decomposition, illustrating the input, goal, tasks, and derived design requirements for each stage.

\subsubsection*{Stage 1: Tumor-Vessel Spatial Analysis}

Taking raw 3D geometric data as input, the objective of this stage is to \textit{discover} the patient-specific anatomy to preliminarily select the target liver segments for resection.
To this end, surgeons need to assemble four categories of spatial evidence through associated tasks (\textbf{T}) before any resection plane can be proposed:

\textbf{T1.1}~\textit{Locate} the tumor and \textit{identify} its spatial extent, specifically determining which liver segments it has invaded;
\textbf{T1.2}~\textit{Compare} the proximity between the tumor boundary and critical vascular structures;
\textbf{T1.3}~\textit{Identify} segment morphological characteristics that govern resection plan and complexity;
\textbf{T1.4}~\textit{Explore} the hepatic vascular architecture to \textit{identify} abnormal branching variants.


\subsubsection*{Stage 2: Interactive Resection Planning}

\revrep{From}{Taking} the candidate target segments \revdel{as input}, \revrep{we aim}{the goal is} to \textit{produce} a plane configuration that satisfies all three critical safety criteria.
Surgeons iteratively adjust plane positions and orientations while verifying \revrep{the}{critical safety} criteria:

\textbf{T2.1}~\textit{Introduce} and \textit{manipulate}  resection planes in 3D anatomical space;
\textbf{T2.2}~\textit{Identify} safety status from system-\textit{derived} FLR, RM, and TSCR safety criteria;
\textbf{T2.3}~\textit{Compare} safety criteria against thresholds and trade \revrep{them off}{off them}.


\begin{figure*}[tb]
  \centering
  \includegraphics[width=\linewidth]{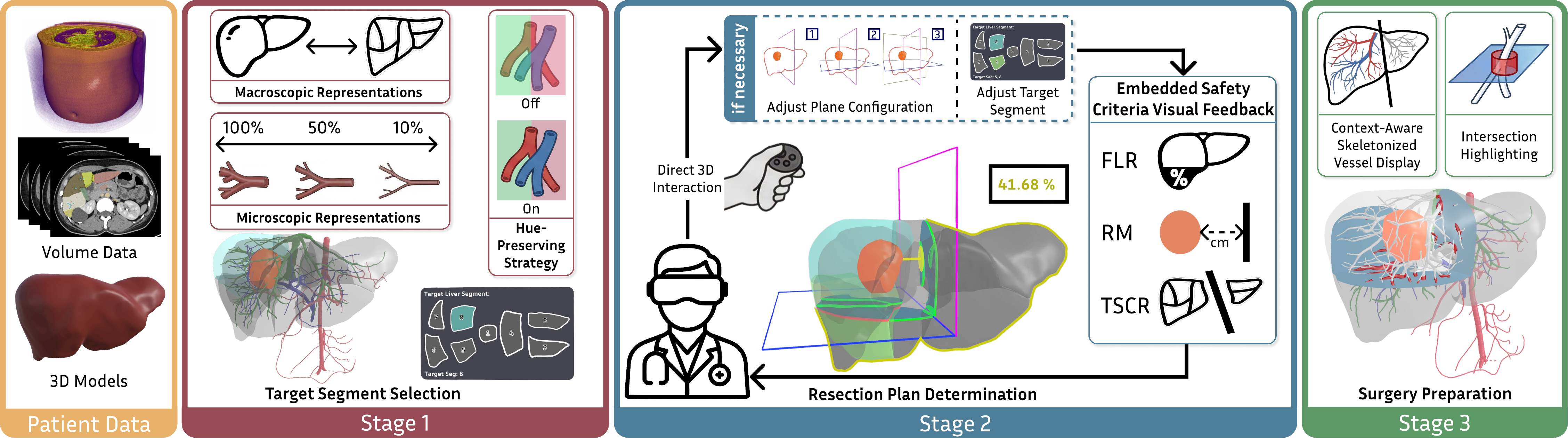}
  \caption{The multi-stage visual analytics workflow in \ourname \revdel{system}. \revdel{In} Stage 1\revrep{:}{,} surgeons confirm the target segments after \revdel{comprehensive} spatial visual analysis\revdel{ of tumor-vessel relationships}. \revdel{In} Stage 2\revrep{:}{,} surgeons finalize the resection \revrep{planes}{strategy} guided by real-time, embedded safety criteria feedback (FLR, RM, and TSCR). \revdel{In} Stage 3\revrep{:}{,} surgeons analyze the spatial relationships between finalized resection planes and vessel branches to build an anticipatory mental model for surgery preparation.}
  \label{fig:multi-stage_workflow}
\end{figure*}

\subsubsection*{Stage 3: Plane-Vessel Intersection Analysis}


With the resection planes finalized, the goal shifts from ``production'' to ``consumption''~\cite{brehmer2013multi}.
Surgeons need to \textit{consume} the completed plan to comprehensively understand the spatial relationships between the intended surgical plan and critical vascular structures:

\textbf{T3.1}~\textit{Identify} which vessel branches fall within the resected versus preserved area;
\textbf{T3.2}~\textit{Locate} the precise spatial intersections between each resection plane and the vessel tree;
\textbf{T3.3}~\textit{Summarize} the distribution and caliber of intersected branches to inform surgery preparation.

If this comprehension process reveals any overlooked issues requiring refinement of the surgical plan, the surgeon may return to Stage 2 for adjustments; otherwise, it concludes the preoperative workflow.


\subsection{Design Goals and Requirements}\label{subsec:goals_and_requirements}

Informed by our clinical observations and task analysis, we identified that the existing \textit{perceptual} and \textit{attention bottlenecks} fundamentally stem from a failure in cognitive offloading~\cite{risko2016cognitive}. 
Specifically, the lack of explicit 3D visual support forces surgeons to rely heavily on internal working memory to mentally reconstruct complex spatial relationships and integrate fragmented safety constraints. 
To alleviate these cognitive burdens, we formulated three primary design goals (\textbf{G}) alongside five actionable design requirements (\textbf{R}).
All requirements aim to externalize these mental computations into interactive visual representations.

\subsubsection*{Goal 1: Externalize Spatial Perception for Anatomical Discovery.}
To support Stage 1 tasks (\textbf{T1.1}--\textbf{T1.4}), the system should facilitate accurate anatomical discovery and spatial comparison by providing clear visual representations.
However, standard alpha blending on 2D screens corrupts the diagnostic color cues (\eg, red arteries, blue veins) that surgeons rely on for vessel identification.
It also degrades the depth perception needed for spatial judgment.
This imposes sustained internal cognitive effort that scales with anatomical complexity, elevating the risk of incorrect segment selection.
To overcome them and achieve \textbf{G1}, we incorporate  two specific requirements in system design:

\textbf{R1:} \emph{Support Context-Preserving Focus.}
To analyze complex intrahepatic relationships (\textbf{T1.1}--\textbf{T1.4}), the system should resolve the severe visual occlusion caused by dense anatomical volumes.
The system should allow surgeons to focus on specific targets such as individual segments or internal vessels, while preserving enough surrounding structural context to maintain comprehensive spatial awareness.

\textbf{R2:} \emph{Maintain Color Fidelity.}
To support reliable tumor and vessel identification (\textbf{T1.1}, \textbf{T1.2}), the visualization should maintain the absolute fidelity of diagnostic color cues even when internal structures are occluded by semi-transparent outer boundaries.
Preserving these hues without blending artifacts ensures that spatial \textit{comparison} tasks are not hindered by color ambiguity.

\subsubsection*{Goal 2: Streamline Multi-Constraint Resection Planning.}
To support interactive resection planning in Stage 2 (\textbf{T2.1}--\textbf{T2.3}), the system should unify 3D resection plane manipulation with real-time, embedded visual feedback of clinical safety constraints. However, current tools rely on 2D control-point interfaces for plane manipulation.
Every adjustment demands clearing and re-specifying control points, making fine-grained iterative refinement impractical.
Furthermore, FLR is checked in a separate panel, while RM and TSCR require manual inspection.
This fragmented workflow forces constant context-switching, disrupting cognitive flow and creating conditions for safety oversight. To overcome these limitations and achieve \textbf{G2}, we incorporate the following two requirements in our system design:

\textbf{R3:} \emph{Prevent Cognitive Disconnect via Direct Spatial Interaction.}
To support the fluid manipulation of resection planes (\textbf{T2.1}), surgeons should be able to interact continuously within the true 3D anatomical space.
This direct manipulation~\cite{shneiderman1983direct} eliminates the cognitive friction caused by using abstract 2D control interfaces for 3D tasks.

\textbf{R4:} \emph{Minimize Attention Shifts via Embedded Visual Feedback.}
To assist surgeons in monitoring critical safety criteria (\textbf{T2.2}, \textbf{T2.3}), the system should tightly couple derived safety criteria with their anatomical referents~\cite{willett2017embedded}.
Encoding quantitative data directly within the primary 3D view eliminates the need to look away at separate panels.

\subsubsection*{Goal 3: Support Anticipatory Mental Model Construction.}
To assist the plane-vessel intersection analysis in Stage 3 (\textbf{T3.1}--\textbf{T3.3}), the system should compute and visualize such intersections, in order to help summarize the vascular context for preparation of the surgery.
However, few existing \revrep{tools provide}{tool provides} dedicated support for this intersection analysis explicitly.
In current practice, surgeons have to mentally reconstruct plane-vessel intersection geometry from 2D cross-sections.
This massive cognitive burden \revdel{can} further scales with the number of planes and anatomical complexity. To overcome these limitations and achieve \textbf{G3}, our system design incorporates the requirement:

\textbf{R5:} \emph{Explicit Plane-Vessel Intersection Visualization.}
To support anticipatory surgical planning (\textbf{T3.1}--\textbf{T3.3}), the system should explicitly show the consequences given by the intended resection plans.
First, the visualization should allow surgeons to clearly \textit{identify} which branches fall within the resected versus preserved areas.
Second, it should clearly mark each plane-vessel intersection at its true anatomical scale, while simultaneously conveying the precise spatial location of these intersections relative to the overall vascular architecture.
\section{\ourname System}

\subsection{Overview}\label{subsec:system_overview}

\begin{figure}[tb]
  \centering
  \includegraphics[width=\columnwidth]{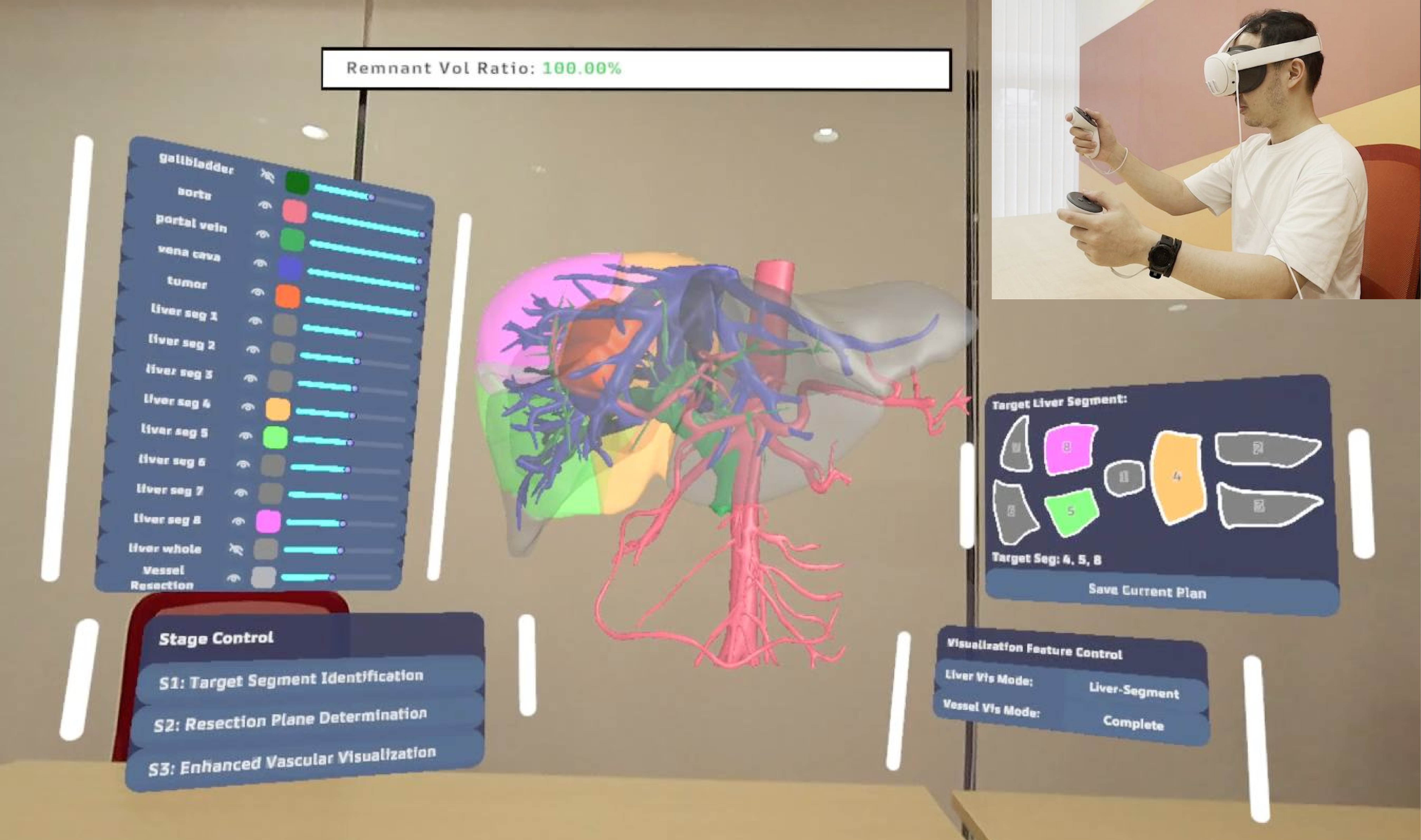}
  \caption{The \ourname interface deployed on a Meta Quest 3 headset.
  The main view shows Stage~1 (tumor-vessel spatial analysis), featuring an integrated information panel for at-a-glance inspection (top), the structure control panel (left), the main anatomical workspace (center), the target segment selection panel (right), and the stage control and visualization feature controls (bottom).
  }
  \label{fig:system_interface}
\end{figure}

Guided by the workflow analysis, design goals and requirements established in \revrep{\cref{sec:design_goals}}{\S\ref{sec:design_goals}}, \ourname provides a stage-adaptive immersive visual analytics framework designed to support hepatobiliary surgeons in ALR surgical planning~(\cref{fig:multi-stage_workflow}). 
Rather than providing a monolithic interface, we decompose the planning workflow into three distinct, sequential stages.
To ensure a fluid analytical experience, the system enforces strict state preservation across stage transitions.


Given the computational constraints of standalone XR headsets, the architecture of \ourname is designed to decouple heavy geometric processing from runtime interactions.
We employ an offline preprocessing pipeline to convert \revrep{reconstructed}{raw patient} anatomical segmentations into optimized 3D surface meshes and analysis-ready spatial data structures.
The interactive client (\cref{fig:system_interface})  loads these surface meshes and precomputed data structures, manages the three-stage workflow execution, and leverages optimized pipelines to ensure real-time surgical planning interactions.
This client is developed using Unity 2022.3 LTS~\cite{unity2022lts} and MRTK3~\cite{MRTK3Unity}, and deployed on a standalone Meta Quest 3 headset.

\begin{figure}[tb]
  \centering
  \includegraphics[width=.98\linewidth]{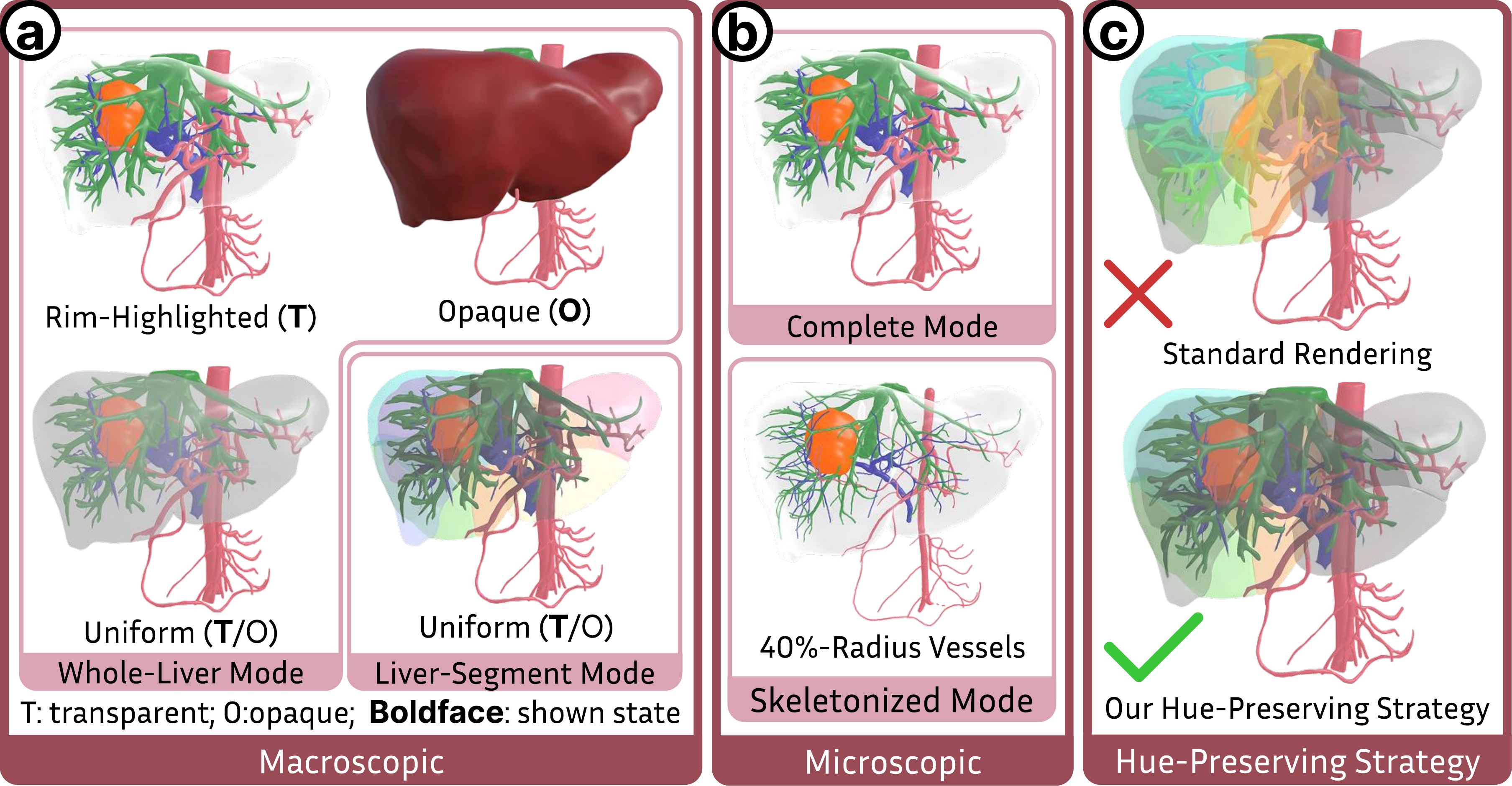}
  \caption{Illustration of the visualization designs in Stage 1.
  (a) Macroscopic Anatomical Representation. A unified whole-liver mode provides three materials—Rim-Highlighted (top left), Opaque (top right), and Uniform (bottom left); an independent liver-segment mode provides the Uniform material (bottom right) for each segment.
  (b) Microscopic Anatomical Representation. Full-radius mode (top) displays the original anatomical vessels. Skeletonized mode (bottom) displays a skeletonized vascular model (\eg, 40\% radius).
  (c) Hue-Preserving Rendering Strategy. Standard rendering (top) distorts internal hues via standard alpha blending; our hue-preserving rendering strategy (bottom) selectively desaturates the external liver segment material where it overlaps internal structures, preserving the original diagnostic colors of vessels and tumors.}
  \label{fig:stage1_visualization}
\end{figure}
\begin{figure}[tb]
  \centering
  \includegraphics[width=0.95\columnwidth]{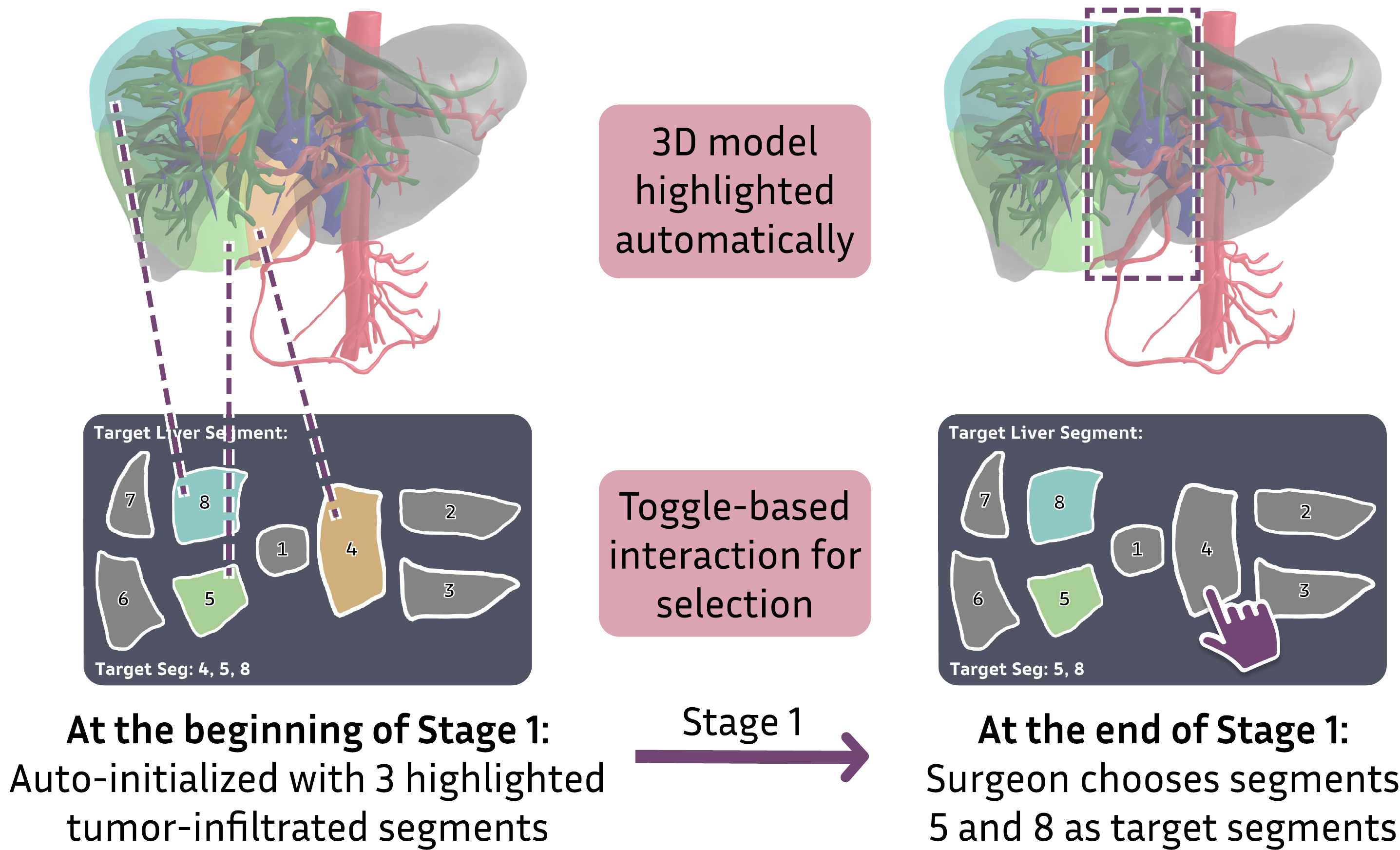}
  \caption{The workflow of Stage 1. \revrep{Left}{The left} column\revrep{:}{ shows} the auto-initialized state, where segments 4, 5, and 8 are highlighted as candidate target segments based on tumor location. \revrep{The}{with the} Target Segment Selection Panel (bottom) corresponding to the 3D model view (top). \revrep{Right}{The right} column\revrep{:}{ shows} the end-of-stage decision after comprehensive spatial visual analysis, where the surgeon finalizes segments 5 and 8 as the target segments.}
  \label{fig:stage1_workflow}
\end{figure}

\subsection{Stage 1: Tumor-Vessel Spatial Analysis}\label{subsec:stage1}

Stage 1 aims to systematically assess tumor-vessel spatial relationships (\textbf{T1.1}--\textbf{T1.4}) to determine the basic target liver segments.
To achieve accurate anatomical discovery (\textbf{G1}) by supporting context-preserving focus (\textbf{R1}) and maintaining color fidelity (\textbf{R2}), we implement the following visual designs.

\subsubsection*{Macroscopic and Microscopic Anatomical Representations}
To support \textbf{R1}, our design is grounded in Focus+Context principles~\cite{hauser2006generalizing}, expanding their scope to both macroscopic and microscopic anatomical scales.
For macroscopic analysis, surgeons can toggle liver visibility between a unified whole-liver mode (context) and an independent liver-segment mode (focus) to assess tumor invasion (\textbf{T1.1}) and efficiently isolate specific segments for inspection (\textbf{T1.3}, \cref{fig:stage1_visualization}(a)).
For microscopic exploration, surgeons can use an interactive slider to seamlessly switch between pre-generated vascular models at 5\% thinning intervals, ranging from full-radius anatomical vessels to highly skeletonized vessels (\cref{fig:stage1_visualization}(b)).
\revrep{Inspired by prior work~\cite{selle2002analysis, eulzer2022vessel}}{Unlike simple global transparency that damages depth perception}, this skeletonized abstraction \revrep{provides a compact overview of}{explicitly preserves} the vascular architecture \revadd{while preserving tumor-vessel spatial relationships}.
This visual encoding actively reduces visual clutter to efficiently expose deep tumor-vessel proximities (\textbf{T1.2}) and rapidly highlight vascular anomalies (\textbf{T1.4}) without removing essential spatial context.
The extraction and generation underlying this topology-aware vascular abstraction are provided in the Supplementary Material.

\subsubsection*{Hue-Preserving Rendering Strategy}
Standard alpha blending often mixes the colors of outer liver surfaces with internal vessels and tumors.
This mixture obscures the normal diagnostic colors (e.g., red for arteries, blue for veins), as demonstrated in \cref{fig:stage1_visualization}(c) top.
To safely maintain color fidelity under occlusion (\textbf{R2}), we adopt a hue-preserving rendering strategy~\cite{chuang2009hue, kuhne2012data} (\cref{fig:stage1_visualization}(c) bottom).
We employ a depth-aware rendering strategy that selectively desaturates the external liver segment material strictly where it overlaps internal structures in the viewer's sight.
Then, internal vessels and tumors consistently retain their diagnostic colors regardless of the outer liver segment's transparency level.
Surgeons can thus reliably identify critical internal structures (\textbf{T1.1}, \textbf{T1.2}) through clear colors.
To achieve this real-time depth-aware rendering on a standalone headset, we implement a Weighted Blended Order-Independent Transparency mechanism~\cite{mcguire2013weighted}, detailed in the Supplementary Material.

\subsubsection*{Target Segment Selection}
To translate the spatial analysis into an preliminary plan, surgeons select the target liver segments via a dedicated panel (\cref{fig:stage1_workflow}).
This interface supports bidirectional toggle interaction, allowing surgeons to explicitly revise the target segments for further planning.

\subsection{Stage 2: Interactive Resection Planning}\label{subsec:stage2}

Stage 2 transitions surgeons from anatomical exploration to active surgical planning, where they determine resection planes (\textbf{T2.1}) while continuously verifying critical safety criteria (\textbf{T2.2}, \textbf{T2.3}).
To streamline this multi-constraint planning process (\textbf{G2}) by preventing cognitive disconnect (\textbf{R3}) and minimizing attention shifts (\textbf{R4}), we implement the following visual designs.

\begin{figure}[tb]
  \centering
  \includegraphics[width=.95\linewidth]{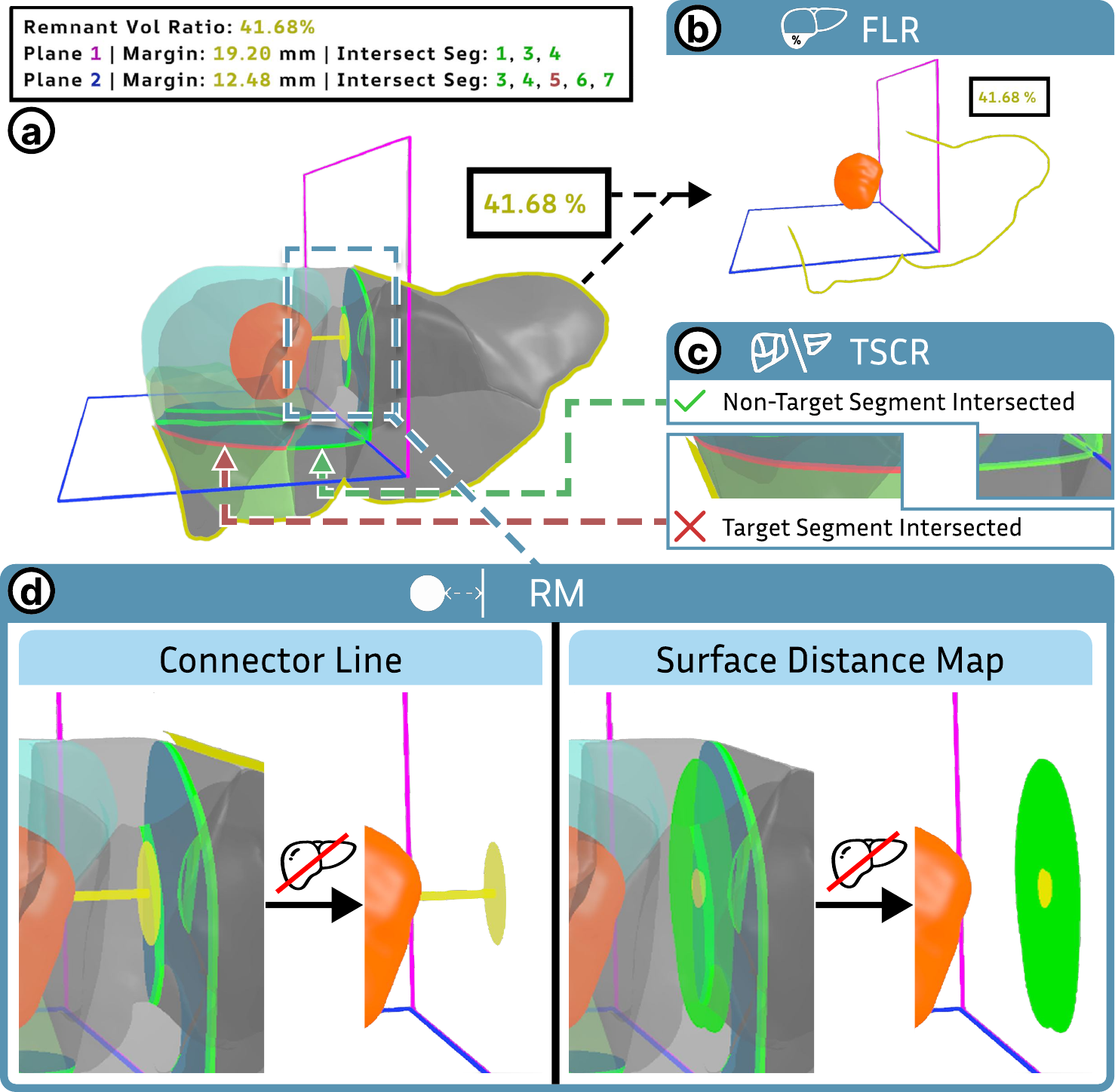}
  \caption{Illustration of the real-time, embedded safety criteria visualization. (a) The integrated information panel presenting a summarized heads-up display (HUD) panel alongside the 3D visualization. (b) Functional Liver Remnant (FLR) feedback showing the current preserved liver volume percentage via a color-coded contour and a floating numerical indicator. (c) Target Segment Complete Removal (TSCR) feedback checking whether the resection plane completely removes target segments, indicated by intersection contours overlaying the liver segment surface. (d) Resection Margin (RM) feedback displayed in complementary modes: connector line mode (left) and surface distance map mode (right), both evaluating the shortest distance between the tumor boundary and the active resection plane.}
  \label{fig:safety_feedback}
\end{figure}

\subsubsection*{Direct 3D Interaction for Plane Manipulation}
To prevent cognitive disconnect via direct spatial interaction (\textbf{R3}), we apply direct manipulation theory~\cite{shneiderman1983direct} to comprehensively bridge the Gulf of Execution~\cite{norman1988design}.
Based on the order of surgery complexity, surgeons first configure the necessary number of resection planes, which the system automatically initializes at anatomically reasonable locations.
Surgeons then grab, move, and rotate these planes directly within the 3D anatomical space using 6-DoF controllers (\textbf{T2.1}).
To realize real-time high-fidelity spatial interactions, we develop a shader-assisted mesh slicing method, detailed in the Supplementary Material.

\subsubsection*{Embedded Visual Feedback for Multi-Constraint Reasoning}
Standard interfaces display safety criteria in disconnected 2D numerical dashboards, forcing surgeons to constantly split their attention between the 3D anatomy and external readout values.
This visual discontinuity fundamentally widens the Gulf of Evaluation.
To explicitly minimize these attention shifts (\textbf{R4}), we utilize the embedded data representations framework~\cite{willett2017embedded} to tightly couple FLR, RM, and TSCR feedback directly onto the focused 3D view (\textbf{T2.2}, \textbf{T2.3}), complemented by an integrated information panel for an overall ``at-a-glance'' plan assessment.

\textit{Functional Liver Remnant (FLR).}
\revrep{We render}{The system renders} a semi-transparent, color-coded outline enclosing the \revdel{preserved} 
remnant liver geometry (\cref{fig:safety_feedback}(a,b).
Alongside this outline, a floating numerical indicator positioned at the liver's upper-right \revrep{offers}{provides} immediate percentage feedback (\cref{fig:safety_feedback}(a,b)).
Both visual elements employ a unified color encoding: green for safe FLR ($>50\%$), yellow for warning ($40$--$50\%$), and red for dangerous ($<40\%$).
By casting the outline directly onto the anatomical surface, surgeons instantly perceive the overall remnant volume safety status through peripheral vision without shifting their gaze away from the focused anatomical region.
To enable \revrep{a}{the} real-time calculation of \revrep{the}{this} enclosed FLR volume against \revadd{the} dynamic planes, we designed a GPU-accelerated voxelization strategy\revrep{; see}{, detailed in} the Supplementary Material.

\textit{Resection Margin (RM).}
To verify sufficient safety margins between the tumor and resection planes, we provide a dual-mode visual encoding.
The \textit{connector line} mode displays the shortest tumor-to-plane distance through a color-coded line with a disk indicator (\cref{fig:safety_feedback}(d) left).
The \textit{surface distance map} mode renders color-coded distance values directly onto the resection plane surface (\cref{fig:safety_feedback}(d) right).
Both modes employ a unified color encoding: green for safe RM ($>2$\,cm), yellow for warning ($1$--$2$\,cm), and red for dangerous ($<1$\,cm).
This dual representation allows surgeons to instantly assess both the local critical proximity and the global safety margin distribution across the entire cut, enabling rapid embedded safety comparisons.
To minimize visual clutter, the RM feedback dynamically activates only while surgeons are actively manipulating the corresponding resection plane.
To evaluate sub-millimeter proximities efficiently during rendering, our backend serializes a dense Signed Distance Field (SDF) 3D texture, optimized as described in the Supplementary Material.

\textit{Target Segment Complete Removal (TSCR).}
Our system highlights plane-segment intersections using dynamic contours overlaying the liver segment surface (\cref{fig:safety_feedback}(c)).
Rather than filling the entire intersected segment with color, we refine the visual encoding to a thin, bright intersection line at the exact boundary.
The contour glows red if the plane intersects a target segment (warning of an incomplete cut) and green otherwise.
Surgeons can easily adjust the plane until no red contours remain, confidently achieving TSCR.

\textit{Integrated Information Panel.}
We provide a centralized heads-up display anchored above the surgical workspace (\cref{fig:safety_feedback}(a), top).
This panel dynamically aggregates all numerical safety criteria simultaneously, including the FLR, per-plane numbers of RM and TSCR.
By utilizing the identical safety color coding as the in-content visual elements, this panel enables rapid ``at-a-glance'' assessment of the overall surgical plan's clinical viability.

\subsection{Stage 3: Plane-Vessel Intersection Analysis}\label{subsec:stage3}

Stage 3 concludes the ALR planning workflow by evaluating the vessel branches \revrep{for}{with respect to} the finalized resection planes (\textbf{T3.1}--\textbf{T3.3}).
To \revrep{construct an}{support} anticipatory mental model \revdel{construction} (\textbf{G3}) based on explicit plane-vessel intersection visualization (\textbf{R5}), we introduce the following visual design.

\begin{figure}[tb]
  \centering
  \includegraphics[width=1.0\linewidth]{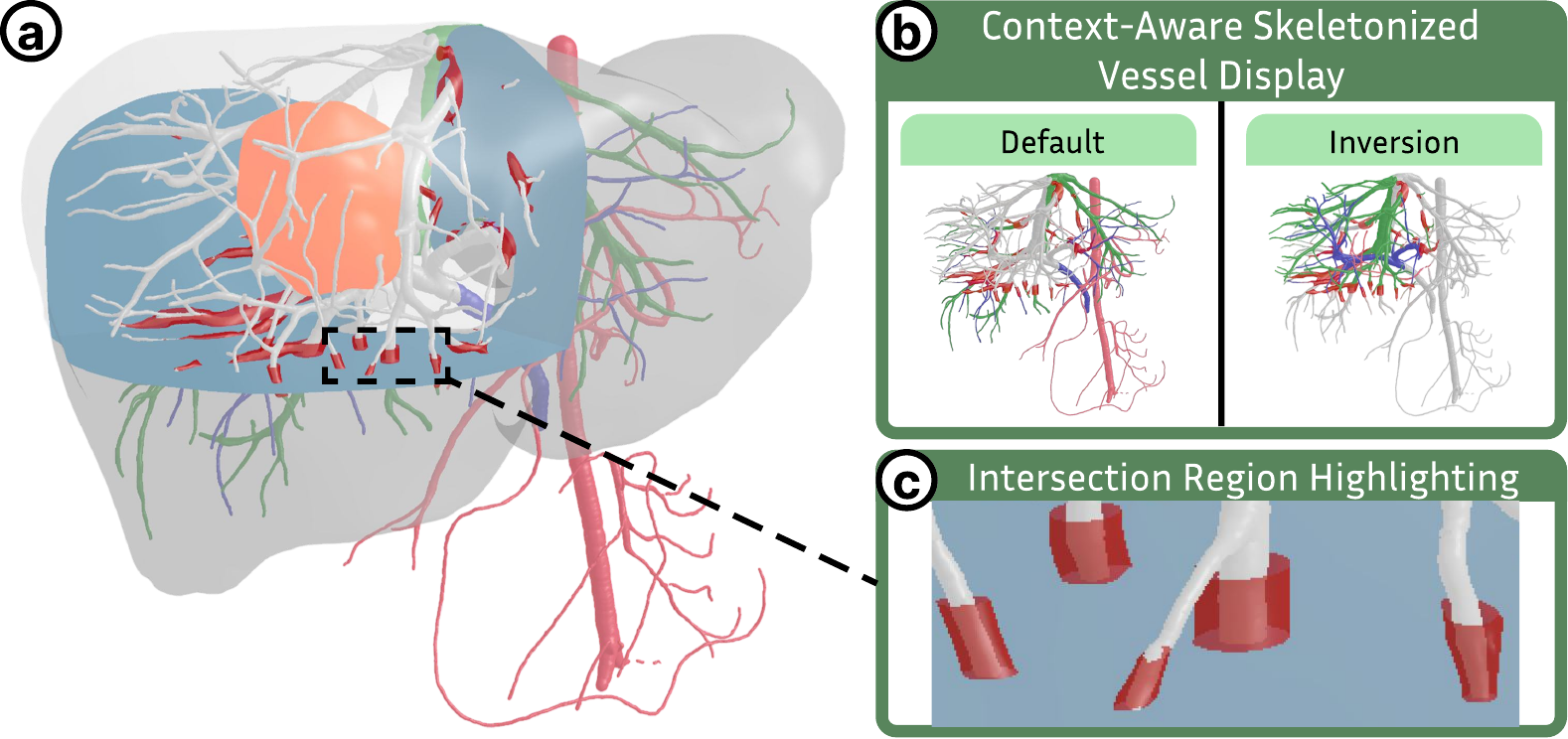}
  \caption{Illustration of visualization designs in Stage 3. (a) The overall visualization shown in Stage 3. (b) Context-aware skeletonized vessel display categorizing vessel branches into resected (desaturated gray) and preserved (diagnostic hues) parts. (c) Zoom-in from (a) showing intersection highlighting, where red collars mark the precise spatial intersections between the resection planes and vessels, explicitly restoring the original anatomical thickness.}
  \label{fig:stage3_visualization}
\end{figure}

\subsubsection*{Focus+Context Plane-Vessel Intersection Visualization}
We implement a structured Focus+Context~\cite{hauser2006generalizing} visual encoding comprising two complementary visual designs.
\textit{Context-Aware Skeletonized Vessel Display}: for the \textit{context}, the system categorizes the skeletonized vessel branches into resected (desaturated gray) and preserved (diagnostic hues) parts, enabling rapid assessment of whether the vessels will be resected or preserved (\textbf{T3.1}, \cref{fig:stage3_visualization}(b)).
\textit{Intersection Highlighting}: for the \textit{focus}, the system renders true-scale translucent red collars at the precise 3D spatial intersections between the resection planes and the vessels (\textbf{T3.2}, \cref{fig:stage3_visualization}(a,c)).
These collars explicitly restore the original anatomical thickness of the intersected vessels, allowing surgeons to quickly estimate the difficulty of the vessel ligation (\textbf{T3.3}) while maintaining full awareness of surrounding vascular topologies.

\section{Evaluation}

We evaluate \ourname through a case study and a within-subject user study with eight hepatobiliary surgeons.
The case study shows how the three-stage workflow supports complete ALR planning in practice.
The expert study then compares \ourname against a desktop baseline to validate the workflow design, the overall planning benefits, and the stage-specific support for \textbf{G1}--\textbf{G3} and \textbf{R1}--\textbf{R5}.

\begin{figure*}[tb]
  \centering
  \includegraphics[width=1.0\linewidth]{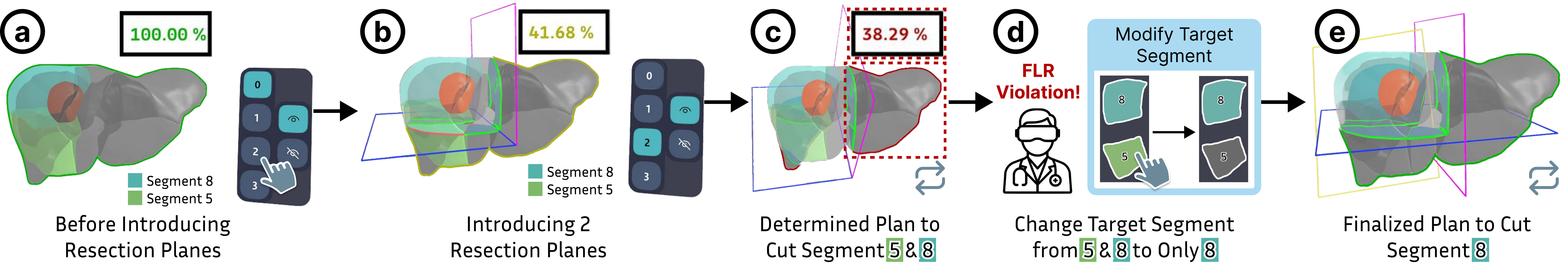}
  \caption{The workflow of Stage 2. (a) Stage-optimized default view emphasizing tumor-segment relationships without active planes. (b) Configuration of the active resection planes, introducing two planes. (c) Manipulating two planes to resect target segments 5 and 8. The embedded visual feedback highlights a Functional Liver Remnant (FLR) violation in red. (d) Reconsidering the surgical strategy due to the FLR violation, the surgeon shifts the target to segment 8 only, transitioning to a three-plane setup. (e) Finalizing the surgical plan by iteratively adjusting the three resection planes, yielding a safe and feasible resection plan that satisfies all safety criteria.}
  \label{fig:stage2_workflow}
\end{figure*}

\subsection{Case Study}

We organized this case study with the resident surgeon S2.
The surgeon used \ourname to plan an ALR surgery for a fully anonymized patient case, which involves a 73\,mm liver tumor primarily located in segment 8 and extending into segments 5 and 4.

S2 begins with Stage 1 to identify the target segments.
The system first highlights segments 4, 5, and 8 in the auto-initialized state (\cref{fig:stage1_workflow}, left).
S2 then inspects tumor extent, tumor-vessel proximity, and segment morphology through context-preserving focus.
By switching between whole-liver and liver-segment modes, and between full and skeletonized vessels, S2 inspects the anatomy from complementary views.
This analysis shows that the tumor extension into segment 4 is moderate and remains close to the segment boundary.
S2 therefore removes segment 4 from the target set and finalizes segments 5 and 8 as the target segments (\cref{fig:stage1_workflow}, right).

S2 then enters Stage 2 to determine the resection plan, introducing two resection planes (\cref{fig:stage2_workflow}(b)) and manipulating them to resect segments 5 and 8 (\cref{fig:stage2_workflow}(c)).
During this process, the embedded visual feedback continuously shows FLR, RM, and TSCR criteria.
The initial plan satisfies RM and TSCR, but the FLR is only 38.29\% (\cref{fig:stage2_workflow}(c)), below the safety threshold.
S2 therefore changes the strategy to a segment-8-centered resection with moderate extension into segments 4 and 5, using three planes (\cref{fig:stage2_workflow}(d)).
After iterative refinement, the final plan satisfies all safety criteria and reaches 52.79\% FLR (\cref{fig:stage2_workflow}(e)).

S2 then proceeds to Stage 3 for plane-vessel intersection analysis (\cref{fig:stage3_visualization}(a)).
The focus+context intersection visualization separates resected and preserved branches, while explicit intersection collars mark the cutting planes (\cref{fig:stage3_visualization}(c)).
This final inspection reveals one portal vein branch and two hepatic vein tributaries that will be encountered during resection, supporting anticipatory mental model construction before surgery.
Overall, this case study shows that \ourname supports a complete and iterative planning process, from preliminary target segment selection to resection planning and final surgical preparation.

\subsection{User Study}

\begin{table}
  \caption{Demographic Information of the participants in user study. The Clinical Experience indicates postgraduate years (PGY) of training for resident surgeons, and attending years for attending surgeons.}
  \label{tab:participants}
  \begin{tabular}{ccccc}
    \toprule
    ID & Gender & Age & Clinical Experience & XR Experience\\
    \midrule
    P1 & Male & 25 & Resident (PGY1) & None\\
    P2 & Male & 31 & Attending (2 years) & Limited\\
    P3 & Male & 27 & Resident (PGY4) & Limited\\
    P4 & Male & 34 & Attending (3 years) & Limited\\
    P5 & Male & 26 & Resident (PGY1) & Limited\\
    P6 & Female & 25 & Resident (PGY1) & Limited\\
    P7 & Male & 29 & Attending (2 years) & None\\
    P8 & Female & 26 & Resident (PGY2) & Familiar\\
    \bottomrule
  \end{tabular}
\end{table}


\emph{Participants.}
We recruited 8 hepatobiliary surgeons from two partner hospitals, including 3 attending surgeons and 5 resident surgeons.
Their mean age was 27.88 years (\emph{SD} = 3.23).
Regarding prior XR experience, 2 participants had none, 5 had limited experience, and 1 was familiar with XR devices.
\cref{tab:participants} summarizes the participant demographics.

\emph{Study Design.}
We conducted a within-subject study that compared \ourname with a desktop baseline.
To control for order effects, half of the participants used \ourname first and the other half used the baseline first.
We selected four clinical cases with two complexity levels, validated by our collaborating surgeons.
The two easier cases involved tumors in segments 2 and 3 without vascular variants.
The two harder cases primarily involved segment 8 and also extended into segments 4 and 5, each with one vascular variant.
The four cases were divided into two case groups, each containing one easier case and one harder case.
Each participant completed one case group on each platform, and the case groups were counterbalanced across platforms.

\emph{Baseline.}
We used 3D Slicer~\cite{fedorov20123d}\revrep{, a widely-used open-source desktop platform, as a representative}{as the} desktop baseline.
\revrep{Based on workflow observations and follow-up interviews, we configured it to reproduce the task-relevant functions of the commercial planning software used by our collaborating surgeons.}{It represents a standard desktop-based ALR planning workflow.}
In our implementation, surgeons defined resection planes with control points, inspected FLR through interface navigation, and measured RM with point-to-point distance tools.
\revrep{This baseline captures a non-immersive workflow that provides the relevant task functionality but lacks the stage-adaptive interaction and embedded spatial feedback introduced in \ourname.}{This baseline captures the indirect interaction and fragmented safety assessment that motivated our design.}

\emph{Tasks.}
For each case, participants completed the full surgical planning workflow from Stage 1 to Stage 3.
To evaluate the workflow in a structured way, we defined four experimental tasks, denoted as E1--E4, that capture the analytical activities.
\begin{itemize}[leftmargin=*, noitemsep, topsep=0pt]
    \item \textbf{E1:} \emph{Tumor Spatial Analysis.}
Participants analyze tumor extent, tumor-vessel relationships, and segment morphology, thereby covering workflow tasks \textbf{T1.1}--\textbf{T1.3}.
\vspace{1mm}
    \item \textbf{E2:} \emph{Vascular Architecture Analysis.}
Participants inspect vascular branching patterns and identify vascular variants, \revdel{thereby} covering \revdel{workflow} tasks \textbf{T1.4}. Together, \textbf{E1} and \textbf{E2} evaluate Stage 1 under \textbf{G1}, \textbf{R1}, and \textbf{R2}.
\vspace{1mm}
    \item \textbf{E3:} \emph{Resection Planning.}
Participants determine a resection plan by introducing and manipulating resection planes while considering FLR, RM, and TSCR, thereby covering workflow tasks \textbf{T2.1}--\textbf{T2.3}.
\vspace{1mm}
    \item \textbf{E4:} \emph{Plane-Vessel Intersection Analysis.}
Participants inspect the finalized plan and identify plane-vessel encounters for surgery preparation, thereby covering workflow tasks \textbf{T3.1}--\textbf{T3.3}.\\
\end{itemize}

\begin{table*}
\centering
\begin{threeparttable}[b]
\caption{Overview of task performance metrics and subjective confidence ratings. }
\label{tab:results_summary}
\begin{tabular}{l|l|l|l|l|l}
\hline
\textbf{\revadd{Scope}} & \textbf{Task/Metric} & \textbf{Our System} & \textbf{Desktop System} & \textbf{Improvement} & \textbf{Sig.} \\
\hline
\multirow{4}{*}{\shortstack[l]{\textbf{\revadd{Overall}}\\\textbf{\revadd{Framework}}\\\textbf{\revadd{(E1-E4 Combined)}}}} & SUS Score $\uparrow$ & $76.25 \pm 13.43$ & $38.44 \pm 16.90$ & $+98.36\%$ & \textbf{***} \\
 & Perceived Workload $\downarrow$ & $21.97 \pm 13.74$ & $32.61 \pm 13.89$ & $-32.63\%$ & \textbf{*} \\
 & Task Completion Time (s) $\downarrow$ & $203.88 \pm 131.84$ & $422.00 \pm 166.38$ & $-51.69\%$ & \textbf{***} \\
 & Confidence Rating $\uparrow$ & $4.22 \pm 0.73$ & $3.84 \pm 0.77$ & $+9.90\%$ & n.s. \\
\hline
\multirow{10}{*}{\shortstack[l]{\textbf{\revadd{Stage 1:}}\\\textbf{\revadd{Tumor-Vessel}}\\\textbf{\revadd{Spatial Analysis}}}} & \textbf{E1: Tumor Spatial Analysis} & & & & \\
 & Spatial Understanding $\uparrow$ & $100\%$ Pass & $100\%$ Pass & No difference & n.a. \\
 & Task Completion Time (s) $\downarrow$ & $32.56 \pm 20.42$ & $74.50 \pm 43.49$ & $-56.30\%$ & \textbf{*} \\
 & Confidence Rating $\uparrow$ & $4.44 \pm 0.56$ & $4.31 \pm 0.65$ & $+2.9\%$ & n.s. \\
\cline{2-6}
 & \textbf{E2: Vascular Architecture Analysis} & & & & \\
 & Variant Detection Rate $\uparrow$ & $100\%$ & $100\%$ & No difference & n.a. \\
 & Task Completion Time (s) $\downarrow$ & $28.94 \pm 29.03$ & $35.81 \pm 28.19$ & $-19.18\%$ & n.s. \\
 & Confidence Rating $\uparrow$ & $4.00 \pm 0.89$ & $4.00 \pm 0.71$ & No difference & n.s. \\
\cline{2-6}
 & \textbf{E1\&E2 Combined} & & & & \\
 & Perceived Workload $\downarrow$ & $21.46 \pm 14.71$ & $32.61 \pm 13.89$ & $-25.95\%$ & n.s. \\
\hline
\multirow{4}{*}{\shortstack[l]{\textbf{\revadd{Stage 2:}}\\\textbf{\revadd{Interactive}}\\\textbf{\revadd{Resection Planning}}}} & \textbf{E3: Resection Planning} & & & & \\
 & Perceived Workload $\downarrow$ & $21.10 \pm 13.58$ & $38.17 \pm 16.11$ & $-44.72\%$ & \textbf{*} \\
 & Task Completion Time (s) $\downarrow$ & $102.50 \pm 67.96$ & $268.25 \pm 117.76$ & $-61.79\%$ & \textbf{***} \\
 & Confidence Rating $\uparrow$ & $4.38 \pm 0.84$ & $3.44 \pm 0.98$ & $+27.33\%$ & \textbf{*} \\
\hline
\multirow{4}{*}{\shortstack[l]{\textbf{\revadd{Stage 3:}}\\\textbf{\revadd{Plane-Vessel}}\\\textbf{\revadd{Intersection Analysis}}}} & \textbf{E4: Plane-Vessel Intersection Analysis} & & & & \\
 & Perceived Workload $\downarrow$ & $23.35 \pm 14.29$ & $30.69 \pm 15.21$ & $-23.92\%$ & n.s. \\
 & Task Completion Time (s) $\downarrow$ & $39.88 \pm 32.45$ & $43.44 \pm 17.29$ & $-8.20\%$ & n.s. \\
 & Confidence Rating $\uparrow$ & $4.06 \pm 1.02$ & $3.62 \pm 0.88$ & $+12.15\%$ & n.s. \\
\hline
\end{tabular}
\begin{tablenotes}
\small
\item[] \noindent
\newcommand{\resultnote}[3]{\parbox[t]{#1}{\hangindent=1em\hangafter=1\textsuperscript{#2}\,#3}}%
\resultnote{.50\linewidth}{1}{\textbf{***} $p < 0.001$, \textbf{**} $p < 0.01$, \textbf{*} $p < 0.05$, n.s. = not significant, n.a. = not applicable.}\hfill
\resultnote{.46\linewidth}{2}{Values shown as Mean±SD.}\par\noindent
\resultnote{.50\linewidth}{3}{$\uparrow$ indicates higher values are better; $\downarrow$ indicates lower values are better.}\hfill
\resultnote{.46\linewidth}{4}{Improvement calculated as percentage change from Desktop to our system.}
\end{tablenotes}
\end{threeparttable}
\end{table*}

\emph{Measures and Analysis.}
We collected task completion time, NASA-TLX workload, confidence rating, and post-study Likert rating for workflow alignment and stage-specific features.
\textbf{E1} included a pass/fail spatial understanding assessment, and \textbf{E2} included a vessel variant detection rate.
After using both systems, participants completed SUS ratings and joined semi-structured interviews.
As the study involved a small number of domain experts, we used a mixed-method analysis.
For quantitative comparisons, we used paired \emph{t}-tests for completion time, workload, and SUS, and Wilcoxon signed-rank tests for confidence ratings.
We reported descriptive statistics and Cohen's \emph{d} effect sizes.
Detailed statistics are provided in the supplementary material.

\subsection{Experimental Results}

\begin{figure}[tb]
  \centering
  \includegraphics[width=\linewidth]{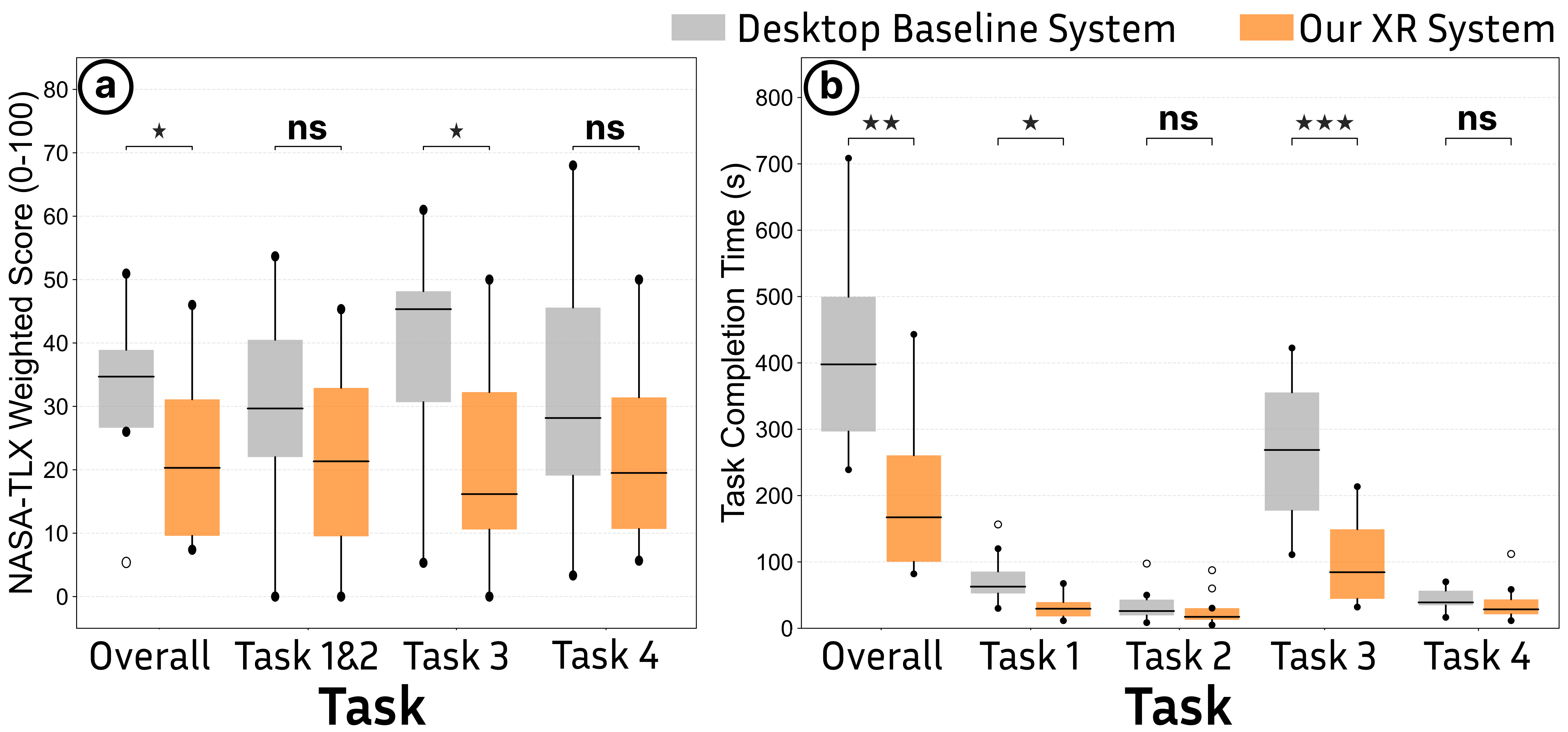}
  \caption{(a) The overall weighted perceived workload (NASA-TLX) score. (b) Task completion time of all experimental tasks. Each boxplot shows the results for each platform on this task, averaged across participants and cases. Symbols $\star (p<0.05)$, $\star\star (p<0.01)$, and $\star\star\star (p<0.001)$ refer to different levels of significance, while \textbf{ns} refers to no significance. Solid lines indicate median values.}
  \label{fig:nasa_tlx-tct}
\end{figure}

\noindent We organize the results from overall workflow effectiveness to stage-specific findings\revrep{:}{.} 
\cref{tab:results_summary} \revrep{on}{summarizes the} quantitative comparisons, \revdel{and} \cref{fig:nasa_tlx-tct} \revrep{on}{shows} workload and completion time distributions, and
\cref{tab:design_evaluation} \revrep{on}{reports} post-study ratings for the stage-specific visual designs and overall workflow validity.
\subsubsection*{Overall Workflow Effectiveness}

\ourname improved the full planning workflow over the desktop baseline.
Across all tasks, it reduced perceived workload by 32.63\% (\emph{d} = 0.92) and task completion time by 51.69\% (\emph{d} = 1.86).
It also improved SUS scores by 98.36\% (\emph{d} = 1.89), while confidence ratings increased by 9.90\% without reaching significance.
These results support the overall utility of the workflow.

The post-study ratings also confirmed strong workflow alignment.
Participants rated Stage 1--3 workflow alignment at 4.88/5, 4.50/5, and 4.88/5, respectively, with cross-stage coherence rated 4.63/5.
These high ratings indicate that our three-stage cognitively-grounded visual analytics workflow aligns well with clinical reasoning.
This interpretation was consistent with the interviews.
P3 remarked that ``\emph{These three stages perfectly mirror how we actually think through surgical planning.}''
In general, these findings show that our visual analytics workflow's system-level gains are not only from the visualization designs, but also from the staged organization of the planning process.

\subsubsection*{Stage 1 Evaluation}

Stage 1 reduced the effort required for anatomical discovery while preserving correctness.
All participants correctly completed both the tumor spatial analysis task and the vascular variant detection task on both platforms.
The main gain therefore lies in efficiency and perceptual clarity rather than raw accuracy.
For \textbf{E1}, \ourname reduced task completion time by 56.30\% (\emph{p} = 0.0215, \emph{d} = 1.04).
For \textbf{E2}, it reduced completion time by 19.18\%, and the combined Stage 1 workload was 25.95\% lower, although these two comparisons were not significant.

The post-study ratings were consistent with this pattern.
Participants gave high ratings to the macroscopic liver analysis modes (4.25/5), the microscopic vessel analysis modes (4.00/5), internal structure visibility clarity (4.50/5), and the target segment selection panel (4.75/5).
In the interviews, participants described the stage as clearer and easier to parse.
P5 noted that ``\textit{The skeletonized vessels made branching patterns easier to inspect without losing context}''.
Overall, these results confirm that the macroscopic and microscopic anatomical representations satisfy \textbf{R1} and the hue-preserving rendering strategy satisfies \textbf{R2}, collectively achieving \textbf{G1} and enabling more efficient spatial evidence assembly.

\subsubsection*{Stage 2 Evaluation}

For \textbf{E3}, \ourname reduced task completion time by 61.79\% (\emph{p} = 0.0006, \emph{d} = 2.07) and perceived workload by 44.72\% (\emph{p} = 0.0107, \emph{d} = 1.22).
It also increased confidence by 27.33\% (\emph{p} = 0.016).
These results show that Stage 2 changes how surgeons perform multi-constraint planning, not only how fast they finish it.

Participants gave high ratings to the direct 3D interaction (4.38/5), FLR embedded visual feedback (4.75/5), RM embedded visual feedback (4.88/5), TSCR embedded visual feedback (4.25/5), and the integrated information panel (4.63/5).
In the interviews, participants clearly attributed the improvement to two factors: direct 3D interaction and embedded critical safety criteria feedback.
P3 described the plane manipulation as ``\emph{very intuitive}''; P4 emphasized that ``\emph{The RM visualization made the critical shortest distance much easier to judge}''; and P8 added that ``\emph{The low interaction cost made them more willing to keep refining the plan instead of stopping early}''.
These findings confirm that direct 3D interaction satisfies \textbf{R3} by eliminating cognitive disconnect, and embedded visual feedback satisfies \textbf{R4} by removing attention shifts, collectively achieving \textbf{G2} and fundamentally improving how surgeons perform multi-constraint planning.

\subsubsection*{Stage 3 Evaluation}

Stage 3 improved \revadd{the} final verification and surgery preparation, \revrep{though}{although} \revrep{being}{the gains were} quantitatively lower than in Stage 2.
For \textbf{E4}, \ourname reduced workload by 23.92\%, reduced task completion time by 8.20\%, and increased confidence by 12.15\%, but none of these differences were significant.
This pattern is expected, \revrep{as}{since} Stage 3 follows Stage 2, during which part of the spatial understanding has already been formed during resection planning.
Its main role is \revrep{thus}{therefore} explicit verification and anticipatory understanding rather than a large standalone speed gain.

In post-study, participants rated the context-aware skeletonized vessel display at 4.25/5 and the intersection highlighting at 4.75/5.
In the interviews, participants P5 and P8 noted that ``\emph{the explicit intersection cues made critical vessel encounters easier to identify before surgery}''.
Particularly, P1, P2, and P7 reported that ``\emph{Stage 3 revealed some details that had been overlooked earlier, and prompted further plan checking}''.
These findings confirm that the Focus+Context plane-vessel intersection visualization satisfies \textbf{R5} by explicitly externalizing spatial relationships, thereby achieving \textbf{G3} and enabling surgeons to build a reliable anticipatory mental model of the finalized plan.
\begin{table}
\begin{threeparttable}[b]
\caption{Post-study ratings of stage-specific visual designs and overall workflow validity for our \ourname system.}
\label{tab:design_evaluation}
\begin{tabular}{l|c}
\hline
\textbf{Visual Design / Workflow Validity} & \textbf{Rating} \\
\hline
\multicolumn{2}{c}{\textbf{Stage 1: Tumor-Vessel Spatial Analysis}} \\
\hline
Macroscopic Liver Analysis Modes & $4.25 \pm 0.89$ \\
Microscopic Vessel Analysis Modes & $4.00 \pm 0.76$ \\
Internal Structure Visibility Clarity & $4.50 \pm 0.76$ \\
Target Segment Selection Panel & $4.75 \pm 0.46$ \\
\hline
\multicolumn{2}{c}{\textbf{Stage 2: Target Segment Selection Panel}} \\
\hline
Direct 3D Interaction & $4.38 \pm 1.19$ \\
Embedded FLR Visual Feedback & $4.75 \pm 0.46$ \\
Embedded RM Visual Feedback & $4.88 \pm 0.35$ \\
Embedded TSCR Visual Feedback & $4.25 \pm 0.71$ \\
Integrated Information Panel & $4.63 \pm 0.52$ \\
\hline
\multicolumn{2}{c}{\textbf{Stage 3: Plane-Vessel Intersection Analysis}} \\
\hline
Context-Aware Skeletonized Vessel Display & $4.25 \pm 1.04$ \\
Intersection Highlighting & $4.75 \pm 0.46$ \\
\hline
\multicolumn{2}{c}{\textbf{Overall Workflow Validity}} \\
\hline
Stage 1 Workflow Alignment & $4.88 \pm 0.35$ \\
Stage 2 Workflow Alignment & $4.50 \pm 0.53$ \\
Stage 3 Workflow Alignment & $4.88 \pm 0.35$ \\
Cross-Stage Logical Coherence & $4.63 \pm 0.52$ \\
Overall System Experience  & $4.75 \pm 0.46$ \\
\hline
\end{tabular}
\begin{tablenotes}
\small
\item [1] All ratings based on 5-point Likert scale questionnaires.
\item [2] Values shown as Mean±SD.
\end{tablenotes}
\end{threeparttable}
\end{table}

\section{Discussion}

\paragraph{Findings}

Beyond the task-level metrics, our evaluation revealed two broader insights relevant to immersive surgical visualization design.

\revadd{\textit{Finding 1: }}\textit{Reduced cognitive burden unlocks deeper planning engagement.}
Surgical planning quality is inherently difficult to quantify, as it depends on surgeon expertise, case complexity, and acceptable trade-offs.
However, a consistent perspective emerged across interviews, with higher-quality plans showing more careful engagement with case-specific details.
Standard desktop tools impose compound operational burdens that discourage iterative refinement.
Reducing perceived workload by 32.63\% and task completion time by 51.69\%, \ourname lowers the operational cost of planning iterations.
Crucially, this does not simply accelerate superficial completion; it shifts the mode of engagement.
As P1 observed: ``\textit{Perhaps because this system provides more intuitive visual feedback, especially the visualization of resection margins, it made my evaluation more detailed, which actually took longer.}''
\revadd{P1's experience was not isolated. In the semi-structured interviews, P4, P7, and P8 expressed similar views, indicating that lower visualization and interaction costs encourage surgeons to further check and refine.}
This suggests that cognitive offloading through embedded visual feedback enables a qualitative shift from merely satisfying constraints to optimizing them, highlighting a distinction with direct clinical implications for plan quality.

\revadd{\textit{Finding 2: }}\textit{Explicit visual representations make spatial reasoning more accessible.}
The macroscopic and microscopic anatomical representations in Stage 1 and the Focus+Context plane-vessel intersection visualization in Stage 3 make complex spatial relationships directly observable rather than mentally reconstructed.
This reduction in perceptual demand was particularly evident among less experienced participants.
As P4 noted: ``\textit{The intuitive visual representation makes it much easier to understand the spatial relationships, which would be incredibly valuable for young surgeons and medical students who are still developing their spatial reasoning skills.}''
This points to a secondary benefit of stage-adaptive visualization beyond supporting expert performance, as it reduces the cognitive barrier for less experienced practitioners to construct \revdel{reliable} surgical spatial models.

\paragraph{Limitation}

\revrep{First,}{The first limitation of our current system is} the planar resection interface, while effective for the majority of clinical cases, \revdel{it} may \revrep{not be sufficient}{be insufficient} for anatomically complex scenarios \revrep{that require}{requiring} irregular resection surfaces.
\revadd{Second, as the XR headset is not weightless, wearing it can be tiring for long planning sessions. Mentally, surgeons need time to get familiar with the device and its controls before use.}
\revrep{Third}{Additionally}, the system's performance depends on the quality of input medical imaging data and anatomical segmentations; poor image quality or segmentation errors may compromise the fidelity of reconstructed 3D models and subsequent planning.
Finally, our evaluation involved eight expert participants from two hospitals, which is appropriate for a domain-expert study but limits broader generalizability and the strength of statistical conclusions.

\paragraph{Future Work}

Current work can be improved in different perspectives.
First, integrating AI-enabled planning assistance (\eg, intelligent suggestions for resection strategies tailored to patient-specific anatomy) can further reduce the cognitive burden of plan synthesis while preserving the surgeon's final decision-making authority.
Second, extending \ourname toward intraoperative use (\eg, combining the preoperative plan with augmented reality overlays during surgical execution) can integrate preoperative intent and intraoperative action.
Moreover, larger multi-institutional studies are needed to validate the clinical utility of the system across diverse surgical contexts and case complexities.
\section{Conclusion}

Surgical planning for anatomical liver resection (ALR) is a complex, multi-stage reasoning process that standard 2D desktop tools fail to adequately support, imposing perceptual and attention bottlenecks.
In this paper, we presented \ournamenospace, a stage-adaptive immersive visual analytics framework for ALR planning, developed through an eight-month longitudinal collaboration with hepatobiliary surgeons.

One of the core contributions of our work is the cognitively grounded characterization of ALR planning as a three-stage sequential process.
This evidence-based decomposition, grounded in \textbf{Multi-Level Typology of Abstract Visualization Tasks}~\cite{brehmer2013multi}, translates clinical observations into three design goals (\textbf{G1}--\textbf{G3}) and five actionable design requirements (\textbf{R1}--\textbf{R5}).
Following this framework, we designed stage-adaptive visualization and interaction techniques to externalize the cognitive demands specific to each stage: macroscopic and microscopic anatomical representations and a hue-preserving rendering strategy for anatomical discovery (\textbf{G1}); direct 3D manipulation with embedded, real-time safety criteria feedback for multi-constraint planning (\textbf{G2}); and a Focus+Context plane-vessel intersection visualization for anticipatory mental model construction (\textbf{G3}).

A within-subject expert study with eight hepatobiliary surgeons \revrep{showed that \ourname improved task efficiency and system usability while reducing cognitive workload relative to the desktop baseline on the controlled planning tasks}{demonstrated that \ourname significantly outperforms a standard desktop baseline}, achieving a 51.69\% reduction in task completion time (Cohen's $d$\,=\,1.86), a 32.63\% reduction in perceived cognitive workload (Cohen's $d$\,=\,0.92), and a 98.36\% improvement in system usability scores (Cohen's $d$\,=\,1.89).
In addition, stage-specific results confirmed that each set of visual designs satisfies its corresponding design requirements and achieves the associated design goal.
Participants also consistently confirmed that the three-stage workflow decomposition aligns with their actual clinical reasoning process.

We believe this work demonstrates a principled approach to designing stage-adaptive surgical visualization systems: by grounding the workflow decomposition in cognitive task analysis and aligning each stage’s visual encoding with its specific perceptual demands, \ourname demonstrates how visual analytics systems can be designed to better support the structured nature of complex clinical reasoning.

\acknowledgments{
  This work was supported by the Research Grants Council of the Hong Kong Special Administrative Region, China, under Project T45-401/22-N; and in part by The Chinese University of Hong Kong, under Projects 4055212 and 4055299.
}

\bibliographystyle{abbrv-doi-hyperref}

\bibliography{main}

@article{andriole2011optimizing,
    author = {Andriole, Katherine P. and Wolfe, Jeremy M. and Khorasani, Ramin and Treves, S. Ted and Getty, David J. and Jacobson, Francine L. and Steigner, Michael L. and Pan, John J. and Sitek, Arkadiusz and Seltzer, Steven E.},
    title = {Optimizing Analysis, Visualization, and Navigation of Large Image Data Sets: One 5000-Section CT Scan Can Ruin Your Whole Day},
    journal = {Radiology},
    volume = {259},
    number = {2},
    pages = {346--362},
    year = {2011},
    doi = {10.1148/radiol.11091276},
    URL = {https://doi.org/10.1148/radiol.11091276}
}

@article{yuan2023extended,
	title = {Extended reality for biomedicine},
	volume = {3},
	issn = {2662-8449},
	url = {https://doi.org/10.1038/s43586-023-00198-y},
	doi = {10.1038/s43586-023-00198-y},
	journal = {Nat. Rev. Methods Primers},
	author = {Yuan, Jie and Hassan, Sohail S. and Wu, Jiaojiao and Koger, Casey R. and Packard, Ren{\'e} R. Sevag and Shi, Feng and Fei, Baowei and Ding, Yichen},
	year = {2023},
    articleno = {14},
    numpages = {16},
}

@article{venkatesan2021virtual,
	title = {Virtual and augmented reality for biomedical applications},
	volume = {2},
	issn = {2666-3791},
	url = {https://doi.org/10.1016/j.xcrm.2021.100348},
	doi = {10.1016/j.xcrm.2021.100348},
	number = {7},
	urldate = {2026-06-09},
	journal = {Cell Rep. Med.},
	publisher = {Elsevier},
	author = {Venkatesan, Mythreye and Mohan, Harini and Ryan, Justin R. and Sch{\"u}rch, Christian M. and Nolan, Garry P. and Frakes, David H. and Coskun, Ahmet F.},
	year = {2021},
    articleno = {100348},
    numpages = {13}
}

@article{boedecker2021using,
	title = {Using virtual {3D}-models in surgical planning: workflow of an immersive virtual reality application in liver surgery},
	volume = {406},
	issn = {1435-2451},
	url = {https://doi.org/10.1007/s00423-021-02127-7},
	doi = {10.1007/s00423-021-02127-7},
	number = {3},
	journal = {Langenbeck's Archives of Surgery},
	author = {Boedecker, Christian and Huettl, Florentine and Saalfeld, Patrick and Paschold, Markus and Kneist, Werner and Baumgart, Janine and Preim, Bernhard and Hansen, Christian and Lang, Hauke and Huber, Tobias},
	year = {2021},
	pages = {911--915},
}

@article{jadhav2022md,
  author={Jadhav, Shreeraj and Kaufman, Arie E.},
  journal={IEEE Trans. Visual Comput. Graphics}, 
  title={MD-Cave: An Immersive Visualization Workbench for Radiologists}, 
  year={2023},
  volume={29},
  number={12},
  pages={4832--4844},
  doi={10.1109/TVCG.2022.3193672},
  url={https://doi.org/10.1109/TVCG.2022.3193672}
}

@article{chheang2021collaborative,
title = {A collaborative virtual reality environment for liver surgery planning},
journal = {Comput. Graphics},
volume = {99},
pages = {234--246},
year = {2021},
doi = {10.1016/j.cag.2021.07.009},
url = {https://doi.org/10.1016/j.cag.2021.07.009},
author = {Vuthea Chheang and Patrick Saalfeld and Fabian Joeres and Christian Boedecker and Tobias Huber and Florentine Huettl and Hauke Lang and Bernhard Preim and Christian Hansen}
}

@article{reitinger2006liver,
  title={Liver surgery planning using virtual reality},
  author={Reitinger, Bernhard and Bornik, Alexander and Beichel, Reinhard and Schmalstieg, Dieter},
  journal={IEEE Comput. Graphics Appl.},
  volume={26},
  number={6},
  pages={36--47},
  year={2006},
  publisher={IEEE},
  doi={10.1109/MCG.2006.131},
  url={https://doi.org/10.1109/MCG.2006.131}
}

@article{hattab2021investigating,
  title={Investigating the utility of VR for spatial understanding in surgical planning: evaluation of head-mounted to desktop display},
  author={Hattab, Georges and Hatzipanayioti, Adamantini and Klimova, Anna and Pfeiffer, Micha and Klausing, Peter and Breucha, Michael and Bechtolsheim, Felix von and Helmert, Jens R and Weitz, J{\"u}rgen and Pannasch, Sebastian and Speidel, Stefanie},
  journal={Sci. Rep.},
  volume={11},
  articleno = {13440},
  numpages = {11},
  year={2021},
  publisher={Nature Publishing Group UK London},
  doi={10.1038/s41598-021-92536-x},
  url={https://doi.org/10.1038/s41598-021-92536-x}
}

@article{reinschluessel2022virtual,
  title={Virtual reality for surgical planning--evaluation based on two liver tumor resections},
  author={Reinschluessel, Anke V and Muender, Thomas and Salzmann, Daniela and Doering, Tanja and Malaka, Rainer and Weyhe, Dirk},
  journal={Frontiers in Surgery},
  volume={9},
  articleno = {821060},
  numpages = {9},
  year={2022},
  publisher={Frontiers Media SA},
  doi={10.3389/fsurg.2022.821060},
  url={https://doi.org/10.3389/fsurg.2022.821060}
}

@article{makuuchi1985ultrasonically,
  title={Ultrasonically guided subsegmentectomy},
  author={Makuuchi, M and Hasegawa, H and Yamazaki, S},
  journal={Surgery, gynecology \& obstetrics},
  volume={161},
  number={4},
  pages={346--350},
  year={1985}
}

@article{hasegawa2005prognostic,
  title={Prognostic impact of anatomic resection for hepatocellular carcinoma},
  author={Hasegawa, Kiyoshi and Kokudo, Norihiro and Imamura, Hiroshi and Matsuyama, Yutaka and Aoki, Taku and Minagawa, Masami and Sano, Keiji and Sugawara, Yasuhiko and Takayama, Tadatoshi and Makuuchi, Masatoshi},
  journal={Annals of surgery},
  volume={242},
  number={2},
  pages={252--259},
  year={2005},
  publisher={LWW},
  doi={10.1097/01.sla.0000171307.37401.db},
  url={https://doi.org/10.1097/01.sla.0000171307.37401.db}
}

@article{nakashima1986pathologic,
  title={Pathologic characteristics of hepatocellular carcinoma},
  author={Nakashima, Toshiro and Kojiro, Masamichi},
  journal={Seminars in liver disease},
  volume={6},
  number={3},
  pages={259--266},
  year={1986},
  organization={Thieme Inc.},
  doi={10.1055/s-2008-1040608},
  url={https://doi.org/10.1055/s-2008-1040608}
}

@article{imamura1999prognostic,
  title={Prognostic significance of anatomical resection and des-$\gamma$-carboxy prothrombin in patients with hepatocellular carcinoma},
  author={Imamura, H and Matsuyama, Y and Miyagawa, Y and Ishida, K and Shimada, R and Miyagawa, S and Makuuchi, M and Kawasaki, S},
  journal={Journal of British Surgery},
  volume={86},
  number={8},
  pages={1032--1038},
  year={1999},
  publisher={Oxford University Press},
  doi={10.1046/j.1365-2168.1999.01185.x},
  url={https://doi.org/10.1046/j.1365-2168.1999.01185.x}
}

@article{huang2017meta,
  title={A Meta-analysis comparing the effect of anatomical resection vs. non-anatomical resection on the long-term outcomes for patients undergoing hepatic resection for hepatocellular carcinoma},
  author={Huang, Xinli and Lu, Sen},
  journal={{HPB}},
  volume={19},
  number={10},
  pages={843--849},
  year={2017},
  publisher={Elsevier},
  doi={10.1016/j.hpb.2017.06.003},
  url={https://doi.org/10.1016/j.hpb.2017.06.003}
}

@article{faria2022liver,
  title={Liver Surgery: Important Considerations for Pre- and Postoperative Imaging},
  author={Faria, Luisa Leit\~{a}o de and Darce, George Felipe and Bordini, Andr\'{e} Leopoldino and Herman, Paulo and Jeismann, Vagner Birk and de Oliveira, Ira\'{\i} Santana and Ortega, Cinthia D. and Rocha, Manoel de Souza},
  journal={{RadioGraphics}},
  volume={42},
  number={3},
  pages={722--740},
  year={2022},
  publisher={Radiological Society of North America},
  doi={10.1148/rg.210124},
  url={https://doi.org/10.1148/rg.210124}
}

@book{couinaud1957foie,
  title={Le foie: {\'e}tudes anatomiques et chirurgicales},
  author={Couinaud, Claude},
  publisher = {Masson},
  address   = {Paris},
  year      = {1957},
  pages     = {530}
}

@article{bray2024global,
  title={Global cancer statistics 2022: {GLOBOCAN} estimates of incidence and mortality worldwide for 36 cancers in 185 countries},
  author={Bray, Freddie and Laversanne, Mathieu and Sung, Hyuna and Ferlay, Jacques and Siegel, Rebecca L and Soerjomataram, Isabelle and Jemal, Ahmedin},
  journal={CA: a cancer journal for clinicians},
  volume={74},
  number={3},
  pages={229--263},
  year={2024},
  publisher={Wiley Online Library},
  doi={10.3322/caac.21834},
  url={https://doi.org/10.3322/caac.21834}
}

@article{chuang2009hue,
  title={Hue-preserving color blending},
  author={Chuang, Johnson and Weiskopf, Daniel and Moller, Torsten},
  journal={IEEE Trans. Visual Comput. Graphics},
  volume={15},
  number={6},
  pages={1275--1282},
  year={2009},
  publisher={IEEE},
  doi={10.1109/TVCG.2009.150},
  url={https://doi.org/10.1109/TVCG.2009.150}
}

@article{kuhne2012data,
  title={A data-driven approach to hue-preserving color-blending},
  author={K{\"u}hne, Lars and Giesen, Joachim and Zhang, Zhiyuan and Ha, Sungsoo and Mueller, Klaus},
  journal={IEEE Trans. Visual Comput. Graphics},
  volume={18},
  number={12},
  pages={2122--2129},
  year={2012},
  publisher={IEEE},
  doi={10.1109/TVCG.2012.186},
  url={https://doi.org/10.1109/TVCG.2012.186}
}

@article{mcguire2013weighted,
  title={Weighted blended order-independent transparency},
  author={McGuire, Morgan and Bavoil, Louis},
  journal={Journal of Computer Graphics Techniques (JCGT)},
  volume={2},
  number={2},
  year={2013},
  pages={122--141}
}

@inproceedings{hauser2006generalizing,
  title={Generalizing Focus+Context Visualization},
  author={Hauser, Helwig},
  booktitle={Scientific Visualization: The Visual Extraction of Knowledge from Data},
  year={2006},
  publisher={Springer},
  address={Berlin, Heidelberg},
  pages={305--327},
  doi={10.1007/3-540-30790-7_18},
  url={https://doi.org/10.1007/3-540-30790-7_18}
}

@article{strasberg2000brisbane,
  title={The {Brisbane} 2000 Terminology of Liver Anatomy and Resections},
  author={Strasberg, SM and Belghiti, J and Clavien, P-A and Gadzijev, E and Garden, JO and Lau, W-Y and Makuuchi, M and Strong, RW},
  journal = {{HPB}},
  volume = {2},
  number = {3},
  pages = {333--339},
  year = {2000},
  publisher={Elsevier},
  doi={10.1016/S1365-182X(17)30755-4},
  url={https://doi.org/10.1016/S1365-182X(17)30755-4}
}

@article{fedorov20123d,
  author={Andriy Fedorov and Reinhard Beichel and Jayashree Kalpathy-Cramer and Julien Finet and Jean-Christophe Fillion-Robin and Sonia Pujol and Christian Bauer and Dominique Jennings and Fiona Fennessy and Milan Sonka and John Buatti and Stephen Aylward and James V. Miller and Steve Pieper and Ron Kikinis},
  title = {{3D Slicer} as an image computing platform for the Quantitative Imaging Network},
  journal = {Magnetic Resonance Imaging},
  volume = {30},
  number = {9},
  pages = {1323--1341},
  year = {2012},
  publisher={Elsevier},
  doi = {10.1016/j.mri.2012.05.001},
  url = {https://doi.org/10.1016/j.mri.2012.05.001}
}

@book{thomas2005illuminating,
  title={Illuminating the Path: The Research and Development Agenda for Visual Analytics},
  editor={Thomas, James J. and Cook, Kristin A.},
  publisher = {{IEEE} Computer Society},
  address   = {Los Alamitos},
  year      = {2005},
  pages     = {186}
}

@inbook{keim2008visual,
  title={Visual analytics: Definition, process, and challenges},
  author={Keim, Daniel and Andrienko, Gennady and Fekete, Jean-Daniel and G{\"o}rg, Carsten and Kohlhammer, J{\"o}rn and Melan{\c{c}}on, Guy},
  booktitle={Information visualization: Human-centered issues and perspectives},
  pages={154--175},
  year={2008},
  address="Berlin, Heidelberg",
  publisher={Springer},
  doi={10.1007/978-3-540-70956-5_7},
  url={https://doi.org/10.1007/978-3-540-70956-5_7}
}

@inbook{militello201315,
  title = {Decision-Centered Design},
  author={Militello, Laura G and Klein, Gary},
  pages={261--271},
  year={2013},
  booktitle = {The Oxford Handbook of Cognitive Engineering},
  publisher = {Oxford University Press},
  doi = {10.1093/oxfordhb/9780199757183.013.0016},
  url = {https://doi.org/10.1093/oxfordhb/9780199757183.013.0016},
}

@article{assadi2022decision,
  title={Decision-centered design of a clinical decision support system for acute management of pediatric congenital heart disease},
  author={Assadi, Azadeh and Laussen, Peter C and Freire, Gabrielle and Ghassemi, Marzyeh and Trbovich, Patricia},
  journal={Frontiers in Digital Health},
  volume={4},
  articleno = {1016522},
  numpages = {11},
  year={2022},
  publisher={Frontiers Media SA},
  doi={10.3389/fdgth.2022.1016522},
  url={https://doi.org/10.3389/fdgth.2022.1016522}
}

@Inbook{Sweller2011,
  author="Sweller, John and Ayres, Paul and Kalyuga, Slava",
  title="The Split-Attention Effect",
  bookTitle="Cognitive Load Theory",
  year="2011",
  publisher="Springer",
  address="New York",
  pages="111--128",
  doi="10.1007/978-1-4419-8126-4_9",
  url="https://doi.org/10.1007/978-1-4419-8126-4_9"
}

@inproceedings{baldonado2000guidelines,
  author = {Wang Baldonado, Michelle Q. and Woodruff, Allison and Kuchinsky, Allan},
  title = {Guidelines for using multiple views in information visualization},
  year = {2000},
  publisher = {ACM},
  address = {New York},
  url = {https://doi.org/10.1145/345513.345271},
  doi = {10.1145/345513.345271},
  booktitle = {Proc. {AVI}},
  pages = {110--119},
  numpages = {10},
  location = {Palermo, Italy}
}

@misc{spillers2004progressive,
  author       = {Spillers, Frank},
  title        = {Progressive Disclosure -- Best Interaction Design Technique?},
  howpublished = {\url{https://frankspillers.com/progressive-disclosure-the-best-interaction-design-technique/}},
  year         = {2004},
  note         = {Accessed: 2026-06-10}
}

@article{wickens2008multiple,
  title={Multiple resources and mental workload},
  author={Wickens, Christopher D},
  journal={Human Factors: The Journal of the Human Factors and Ergonomics Society},
  volume={50},
  number={3},
  pages={449--455},
  year={2008},
  publisher={SAGE Publications Sage CA: Los Angeles, CA},
  doi={10.1518/001872008X288394},
  url={https://doi.org/10.1518/001872008X288394}
}

@article{sedlmair2012design,
  author={Sedlmair, Michael and Meyer, Miriah and Munzner, Tamara},
  journal={IEEE Trans. Visual Comput. Graphics}, 
  title={Design Study Methodology: Reflections from the Trenches and the Stacks}, 
  year={2012},
  volume={18},
  number={12},
  pages={2431--2440},
  doi={10.1109/TVCG.2012.213},
  url={https://doi.org/10.1109/TVCG.2012.213}
}

@inproceedings{bhat2023towards,
author = {Bhat, Karthik S. and Kumar, Neha and Shamanna, Karthik and Kwatra, Nipun and Jain, Mohit},
title = {Towards Intermediated Workflows for Hybrid Telemedicine},
year = {2023},
publisher = {ACM},
address = {New York},
url = {https://doi.org/10.1145/3544548.3580653},
doi = {10.1145/3544548.3580653},
booktitle = {Proc. {CHI}},
articleno = {347},
numpages = {17},
location = {Hamburg, Germany}
}

@article{shao2020teaching,
  author = {Shao, Qijia and Sniffen, Amy and Blanchet, Julien and Hillis, Megan E. and Shi, Xinyu and Haris, Themistoklis K. and Liu, Jason and Lamberton, Jason and Malzkuhn, Melissa and Quandt, Lorna C. and Mahoney, James and Kraemer, David J. M. and Zhou, Xia and Balkcom, Devin},
  title = {Teaching {American} Sign Language in Mixed Reality},
  year = {2020},
  issue_date = {December 2020},
  publisher = {ACM},
  address = {New York},
  volume = {4},
  number = {4},
  url = {https://doi.org/10.1145/3432211},
  doi = {10.1145/3432211},
  journal = {Proc. ACM Interact. Mobile Wearable Ubiquitous Technol.},
  articleno = {152},
  numpages = {27}
}

@article{galle2018easl,
  title={EASL Clinical Practice Guidelines: Management of hepatocellular carcinoma},
  author={Galle, Peter R and Forner, Alejandro and Llovet, Josep M and Mazzaferro, Vincenzo and Piscaglia, Fabio and Raoul, Jean-Luc and Schirmacher, Peter and Vilgrain, Val{\'e}rie},
  journal={Journal of Hepatology},
  volume={69},
  number={1},
  pages={182--236},
  year={2018},
  publisher={Elsevier},
  doi={10.1016/j.jhep.2018.03.019},
  url={https://doi.org/10.1016/j.jhep.2018.03.019}
}

@article{memeo2021optimization,
  author = {Riccardo Memeo and Maria Conticchio and Emmanuel Deshayes and Silvio Nadalin and Astrid Herrero and Boris Guiu and Fabrizio Panaro},
  title = {Optimization of the future remnant liver: review of the current strategies in {Europe}},
  journal = {HepatoBiliary Surgery and Nutrition},
  volume = {10},
  number = {3},
  pages = {350--363},
  year = {2021},
  doi={10.21037/hbsn-20-394},
  url = {https://doi.org/10.21037/hbsn-20-394}
}

@article{lin2022prognostic,
  title={Prognostic Impact of Surgical Margin in Hepatectomy on Patients With Hepatocellular Carcinoma: A Meta-Analysis of Observational Studies},
  author={Lin, Yeting and Xu, Jiaxuan and Hong, Jiaze and Si, Yuexiu and He, Yujing and Zhang, Jinhang},
  journal={Frontiers in Surgery},
  volume={9},
  articleno = {810479},
  numpages = {11},
  year={2022},
  publisher={Frontiers Media SA},
  doi={10.3389/fsurg.2022.810479},
  url={https://doi.org/10.3389/fsurg.2022.810479}
}

@misc{MRTK3Unity,
  author       = {{Mixed Reality Toolkit Organization}},
  title        = {{MRTK3}: Mixed Reality Toolkit for {Unity}},
  year         = {2023},
  howpublished = {\url{https://github.com/MixedRealityToolkit/MixedRealityToolkit-Unity}},
  note         = {Accessed 2026-03-08}
}

@article{brehmer2013multi,
  author={Brehmer, Matthew and Munzner, Tamara},
  journal={IEEE Trans. Visual Comput. Graphics},
  title={A Multi-Level Typology of Abstract Visualization Tasks}, 
  year={2013},
  volume={19},
  number={12},
  pages={2376--2385},
  doi={10.1109/TVCG.2013.124},
  url={https://doi.org/10.1109/TVCG.2013.124}
}

@article{risko2016cognitive,
  title = {Cognitive Offloading},
  journal = {Trends in Cognitive Sciences},
  volume = {20},
  number = {9},
  pages = {676--688},
  year = {2016},
  issn = {1364-6613},
  doi = {10.1016/j.tics.2016.07.002},
  url = {https://doi.org/10.1016/j.tics.2016.07.002},
  author = {Evan F. Risko and Sam J. Gilbert}
}

@article{knittel2025embryoprofiler,
  author={Knittel, Johannes and Warchol, Simon and Troidl, Jakob and Brumar, Camelia D. and Yang, Helen Yu and M{\"o}rth, Eric and Kr{\"u}ger, Robert and Needleman, Daniel and Ben-Yosef, Dalit and Pfister, Hanspeter},
  journal={IEEE Trans. Visual Comput. Graphics}, 
  title={EmbryoProfiler: A Visual Clinical Decision Support System for IVF}, 
  year={2026},
  volume={32},
  number={1},
  pages={1262--1272},
  doi={10.1109/TVCG.2025.3634780},
  url={https://doi.org/10.1109/TVCG.2025.3634780}
}

@misc{materialise_mimics,
  author       = {{Materialise NV}},
  title        = {Mimics Innovation Suite},
  year         = {2023},
  howpublished = {\url{https://www.materialise.com/en/healthcare/mimics}},
  note         = {Accessed: 2026-06-10}
}

@ARTICLE{willett2017embedded,
  author={Willett, Wesley and Jansen, Yvonne and Dragicevic, Pierre},
  journal={IEEE Trans. Visual Comput. Graphics}, 
  title={Embedded Data Representations}, 
  year={2017},
  volume={23},
  number={1},
  pages={461--470},
  doi={10.1109/TVCG.2016.2598608},
  url={https://doi.org/10.1109/TVCG.2016.2598608}
}

@article{horeman2012visual,
  title={Visual force feedback in laparoscopic training},
  author={Horeman, Tim and Rodrigues, Sharon P and van den Dobbelsteen, John J and Jansen, Frank-Willem and Dankelman, Jenny},
  journal={Surgical Endoscopy},
  volume={26},
  pages={242--248},
  year={2012},
  publisher={Springer},
  doi={10.1007/s00464-011-1861-4},
  url={https://doi.org/10.1007/s00464-011-1861-4}
}

@article{tu2024head,
  author={Tu, Mingxiao and Jung, Hoijoon and Kim, Jinman and Kyme, Andre},
  journal={IEEE J. Biomed. Health. Inf.}, 
  title={Head-Mounted Displays in Context-Aware Systems for Open Surgery: A State-of-the-Art Review}, 
  year={2025},
  volume={29},
  number={2},
  pages={1165--1175},
  doi={10.1109/JBHI.2024.3485023},
  url={https://doi.org/10.1109/JBHI.2024.3485023}
}

@inproceedings{tatzgern2014hedgehog,
  author={Tatzgern, Markus and Kalkofen, Denis and Grasset, Raphael and Schmalstieg, Dieter},
  booktitle={Proc. IEEE VR}, 
  title={Hedgehog labeling: View management techniques for external labels in {3D} space}, 
  year={2014},
  pages={27--32},
  address={New York},
  publisher={{IEEE}},
  doi={10.1109/VR.2014.6802046},
  url={https://doi.org/10.1109/VR.2014.6802046}
}

@article{lin2021labeling,
  author={Lin, Tica and Yang, Yalong and Beyer, Johanna and Pfister, Hanspeter},
  journal={IEEE Trans. Visual Comput. Graphics}, 
  title={Labeling Out-of-View Objects in Immersive Analytics to Support Situated Visual Searching}, 
  year={2023},
  volume={29},
  number={3},
  pages={1831--1844},
  doi={10.1109/TVCG.2021.3133511},
  url={https://doi.org/10.1109/TVCG.2021.3133511}
}

@book{marriott2018immersive,
  editor={Marriott, Kim and Schreiber, Falk and Dwyer, Tim and Klein, Karsten and Riche, Nathalie Henry and Itoh, Takayuki and Stuerzlinger, Wolfgang and Thomas, Bruce H.},
  title     = {Immersive Analytics},
  series    = {Lecture Notes in Computer Science},
  volume    = {11190},
  publisher = {Springer},
  address   = {Cham},
  year      = {2018},
  doi       = {10.1007/978-3-030-01388-2},
  url={https://doi.org/10.1007/978-3-030-01388-2}
}

@book{norman1988design,
  author    = {Don Norman},
  title     = {The Design of Everyday Things},
  edition   = {Revised and Expanded},
  publisher = {Basic Books},
  address   = {New York},
  year      = {2013},
  isbn      = {978-0-465-05065-9}
}

@article{shneiderman1983direct,
  author={Shneiderman, Ben},
  journal={Computer}, 
  title={Direct Manipulation: A Step Beyond Programming Languages}, 
  year={1983},
  volume={16},
  number={8},
  pages={57--69},
  doi={10.1109/MC.1983.1654471}
}

@misc{unity2022lts,
  author       = {{Unity Technologies}},
  title        = {{Unity} 2022.3 {LTS}},
  year         = {2023},
  howpublished = {\url{https://unity.com/}},
  note         = {Accessed: 2026-06-10}
}

@article{fonnet2019survey,
  author={Fonnet, Adrien and Pri{\'e}, Yannick},
  journal={IEEE Trans. Visual Comput. Graphics}, 
  title={Survey of Immersive Analytics}, 
  year={2021},
  volume={27},
  number={3},
  pages={2101--2122},
  doi={10.1109/TVCG.2019.2929033}
}

@article{chandler1992split,
  author = {Chandler, Paul and Sweller, John},
  title = {THE SPLIT-ATTENTION EFFECT AS A FACTOR IN THE DESIGN OF INSTRUCTION},
  journal = {British Journal of Educational Psychology},
  volume = {62},
  number = {2},
  pages = {233--246},
  doi = {10.1111/j.2044-8279.1992.tb01017.x},
  year = {1992}
}

@inproceedings{ens2021grand,
  author = {Ens, Barrett and Bach, Benjamin and Cordeil, Maxime and Engelke, Ulrich and Serrano, Marcos and Willett, Wesley and Prouzeau, Arnaud and Anthes, Christoph and B\"{u}schel, Wolfgang and Dunne, Cody and Dwyer, Tim and Grubert, Jens and Haga, Jason H. and Kirshenbaum, Nurit and Kobayashi, Dylan and Lin, Tica and Olaosebikan, Monsurat and Pointecker, Fabian and Saffo, David and Saquib, Nazmus and Schmalstieg, Dieter and Szafir, Danielle Albers and Whitlock, Matt and Yang, Yalong},
  title = {Grand Challenges in Immersive Analytics},
  year = {2021},
  isbn = {9781450380966},
  publisher = {ACM},
  address = {New York},
  url = {https://doi.org/10.1145/3411764.3446866},
  doi = {10.1145/3411764.3446866},
  booktitle = {Proc. {CHI}},
  articleno = {459},
  numpages = {17},
  location = {Yokohama, Japan}
}

@article{quinn2022three,
  title = {The three ghosts of medical AI: Can the black-box present deliver?},
  journal = {Artif. Intell. Med.},
  volume = {124},
  articleno = {102158},
  numpages = {8},
  year = {2022},
  issn = {0933-3657},
  doi = {10.1016/j.artmed.2021.102158},
  author = {Thomas P. Quinn and Stephan Jacobs and Manisha Senadeera and Vuong Le and Simon Coghlan},
}

@article{bismuth1982surgical,
  author = {Bismuth, Henri},
  title = {Surgical anatomy and anatomical surgery of the liver},
  journal = {World Journal of Surgery},
  volume = {6},
  number = {1},
  pages = {3--9},
  doi = {10.1007/BF01656368},
  year = {1982}
}

@article{stolper2014progressive,
  author={Stolper, Charles D. and Perer, Adam and Gotz, David},
  journal={IEEE Trans. Visual Comput. Graphics}, 
  title={Progressive Visual Analytics: User-Driven Visual Exploration of In-Progress Analytics}, 
  year={2014},
  volume={20},
  number={12},
  pages={1653--1662},
  doi={10.1109/TVCG.2014.2346574}
}

@article{selle2002analysis,
  author={Selle, D. and Preim, B. and Schenk, A. and Peitgen, H.-O.},
  journal={IEEE Trans. Med. Imaging}, 
  title={Analysis of vasculature for liver surgical planning}, 
  year={2002},
  volume={21},
  number={11},
  pages={1344--1357},
  doi={10.1109/TMI.2002.801166}
}

@article{eulzer2022vessel,
author = {Eulzer, P. and Meuschke, M. and Mistelbauer, G. and Lawonn, K.},
title = {Vessel Maps: A Survey of Map-Like Visualizations of the Cardiovascular System},
journal = {Comput. Graphics Forum},
volume = {41},
number = {3},
pages = {645--673},
doi = {10.1111/cgf.14576},
year = {2022}
}

\end{document}